\documentclass[12pt]{article}

\usepackage[english]{babel}
\usepackage[cp1251]{inputenc}
\usepackage[T1]{fontenc}

\pdfoutput=1
\usepackage{makeidx}
\usepackage{amssymb}
\usepackage{amsfonts}
\usepackage{amsmath}
\usepackage{graphicx}
\usepackage{setspace}
\usepackage{authblk}
\usepackage{units}
\usepackage{appendix}
\usepackage{xparse}
\usepackage{cite}
\usepackage{hyperref}
\usepackage[left=2.1cm,top=2.5cm,right=2.1cm]{geometry}

\begin{document}

\title{Electromagnetic radiation of accelerated charged particle in the
	framework of a semiclassical approach}
\author{T. C. Adorno$^{1}$\thanks{Tiago.Adorno@xjtlu.edu.cn},
	    A. I. Breev$^{2}$\thanks{breev@mail.tsu.ru}, 
	    A. J. D. Farias Jr$^{3}$\thanks{a.jorgedantas@gmail.com},
		D. M. Gitman$^{2,4,5}$\thanks{dmitrygitman@hotmail.com}\\
		$^{1}$ Department of Physics, Xi'an Jiaotong-Liverpool University, 111
		Ren'ai Road,\\
		Suzhou Dushu Lake Science and Education Innovation District,
		Suzhou Industrial Park, Suzhou 215123, People's Republic of China;\\
	    $^{2}$ Department of Physics, Tomsk State University, \\
	    Lenin Ave. 36, 634050, Tomsk, Russia;\\
	    $^{3}$ Centro de Ci\^{e}ncias Exatas e da Natureza, Universidade Federal da
		Para\'{i}ba, Cidade Universit\'{a}ria, CEP 58051-090, Jo\~{a}o Pessoa, Brazil;\\
	    $^{4}$ P.N. Lebedev Physical Institute, 53 Leninskiy prospect, 119991
		Moscow, Russia;\\
	    $^{5}$ Institute of Physics, University of S\~{a}o Paulo,\\ 
	    Rua do Mat\~{a}o, 1371, CEP 05508-090, S\~{a}o Paulo, SP, Brazil.}    
	    
\maketitle

\abstract{
	We address the problem of the electromagnetic radiation produced by charge
	distributions in the framework of a semiclassical approach proposed in the
	work by Bagrov, Gitman, Shishmarev and Farias [J. Synchrotron Rad. (2020).
	\textbf{27}, 902--911]. In this approach, currents, generating the radiation
	are considered classically, while the quantum nature of the radiation is kept
	exactly. Quantum states of the electromagnetic field are solutions of
	Schr\"{o}dinger's equation and relevant quantities to the problem are evaluated
	with the aid of transition probabilities. This construction allows us to
	introduce the quantum transition time in physical quantities and assess its
	role in radiation problems by classical currents. We study radiated
	electromagnetic energies in detail and present a definition for the rate at
	which radiation is emitted from sources. In calculating the total energy and
	rate radiated by a pointlike charged particle accelerated by a constant and
	uniform electric field, we discover that our results are compatible with
	results obtained by other authors in the framework of the classical radiation
	theory under an appropriate limit. We also perform numerical and asymptotic
	analysis of the results.
	
}



\section{Introduction\label{S1}}

It is known that a change in the state of motion of charged particles is
usually accompanied by an electromagnetic radiation. The most prominent
examples are the radiation of accelerated charges undergoing rectilinear
motion and of charges performing circular motion. The most adequate and
accurate description of the process should be carried out in the framework of
$QED$; see e.g. Refs. \cite{Heitl36,Schwe61,BogSh80}. However, in a
number of cases, specific technical difficulties arising from the direct
application of the $QED$ formalism can be avoided resorting to approximations,
which often allows one quite accurately unveiling the essence of the physical
effect. Here, the so-called classical description of the effect is most often
used, which is as follows: the motion of charged particles is considered
within the framework of classical (non-relativistic or relativistic)
mechanics, then using Maxwell's equations, the electromagnetic field generated
by this motion is restored (for example, in the form of Li\'{e}nard-Wiechert
potentials), finally, the assumption is made (which corresponds to an
approximation) that the observed electromagnetic radiation generated by the
motion of the charges must be calculated as the energy flux determined by the
Umov-Poynting vector, see e.g. Refs. \cite{LanLi71,Jacks99,SokTe68}. For
example, in the framework of the classical description, a formula for the
angular distribution of the synchrotron radiation power was obtained by Schott
\cite{Schott}. An alternative derivation of this result and its deep analysis,
especially for high-energy relativistic electrons, was given by Schwinger
\cite{Schwi49}. The essence of quantum corrections to the classical result was
first pointed out in Ref. \cite{Schwi54}. $QED$ calculation of such
corrections was presented in Ref. \cite{SokTe68} within the Furry picture
\cite{Furry} (based on exact solutions of the Dirac equations with a magnetic
field). In the framework of the source theory \cite{Schwi}, Schwinger had
presented an original derivation of similar results \cite{Schwi73}.

The electromagnetic radiation produced by a uniformly
accelerated\footnote{Following the usual convent, by \textquotedblleft uniform
	acceleration\textquotedblright\ we mean a time-independent homogeneous
	acceleration in the particle instantaneous rest frame.} charged particle was
from the beginning a subject of theoretical debate. The central aspect of the
discussion was the question of whether such a particle could radiate at all or
not. One of the arguments contrary to the radiation was raised by Pauli
\cite{Pauli58}, who calculated the electromagnetic fields produced by the
particle and concluded that such a particle could not radiate since the
corresponding fields did not contain wave-like contributions. In order to
reach this conclusion, Pauli used results obtained earlier by Born
\cite{Born1909} and elements of conformal invariance of Maxwell's equations.
On the other hand, years earlier, Schott \cite{Schott} had calculated the
electromagnetic fields radiated by the accelerated particle and computed the
energy rate through the Poynting's theorem. He discovered that the energy
emission rate is proportional to the square of the particle's acceleration and
mentioned that his result was in agreement with the Larmor's classical result
\cite{Larmor1987}, currently known as the Larmor's formula \cite{Jacks99}. The
most compelling arguments in favor of the radiation were later given by Fulton
and Rohrlich in Refs. \cite{FulRoh60,Rohrlich61,Rohrlich63,RohrlichBook}. In
these works, the authors presented a thorough discussion of the relevant
publications and calculated the energy and momentum emission rate in a
covariant form. Few years later, Nikishov and Ritus, using Landau-Lifshitz
formula for the spectral distribution of the emitted electromagnetic energy
\cite{LanLi71}, had calculated the energy spectrum and the energy emission
rate in a covariant form \cite{NikRit69}. More recently, Eriksen and Gron
extensively discussed the problem in a series of papers
\cite{EriGro00-I,EriGro00-II,EriGro00-III,EriGro00-IV,EriGro00-V} and
presented a study of radiation reaction effects. At present, it is widely
accepted that uniformly accelerated particles radiate, and the classical
derivation of its energy rate is presented in many textbooks; see e.g. Refs.
\cite{LanLi71,Jacks99,SchRMT-book,ItzZub,Barut}.

In the work \cite{BagGiSF20}, it was proposed the so-called semiclassical
approach for describing quantum properties of radiation of currents of charged
particles. In this approach, currents, generating the radiation are considered
classically, whereas the quantum nature of the radiation is taken into account
exactly. Universal formulas describing multiphoton radiation were derived. The
approach does not require knowledge of the exact solutions of relativistic
wave equations with external fields; hence technical difficulties associated
with the use of the Furry picture do not arise. Moreover, the semiclassical
approach can be applied to any trajectory performed by the particle, even
including cases with backreaction, as these can be accounted for by solving
the Lorentz equations with radiation-reaction terms. We note\textit{ }that in
the framework of the semiclassical approach one can directly calculate the
radiation emitted from any trajectory of a charged particle, whereas, in QED,
the technics of calculating photon transition amplitudes (let say in the Furry
picture) is adopted only for charge motions caused by external electromagnetic
fields. However, an univocal correspondence between every charge trajectory
and a corresponding electromagnetic field does not exist. The efficacy of the
semiclassical approach was demonstrated in calculating synchrotron
\cite{BagGiSF20}\ and undulator \cite{ShiLeBG21} radiations. In the present
article we apply the semiclassical approach to study electromagnetic radiation
by uniformly accelerated charged particles. In Sect. \ref{S2} we present the
basic formulation of the semiclassical description of electromagnetic
radiation in the general case, refining and supplementing the results
presented in the original Refs. \cite{BagGiSF20,ShiLeBG21}. In Sect. \ref{S3},
we apply these results to calculate electromagnetic energies and rates
radiated by an accelerated charge. In Sect. \ref{S4}, we compare the obtained
results with the above cited calculations by other authors.

\section{Semiclassical description of electromagnetic radiation induced by
	classical currents\label{S2}}

\subsection{Photon state vector\label{S2.1}}

The semiclassical approach under consideration is based on the possibility of
an accurate analytical construction of the state vector $\left\vert
\Psi\left(  t\right)  \right\rangle $ of the quantized electromagnetic field
interacting with a classical current. We describe below details on the
construction of such a state vector and its representation in a form
convenient for our purposes. The state vector $\left\vert \Psi\left(
t\right)  \right\rangle $ satisfies the Schr\"{o}dinger equation and an initial
condition $\left\vert \Psi\right\rangle _{\mathrm{in}}$ at the initial time
instant $t_{\mathrm{in}}$,%
\begin{equation}
	i\hslash\partial_{t}\left\vert \Psi\left(  t\right)  \right\rangle =\hat
	{H}\left(  t\right)  \left\vert \Psi\left(  t\right)  \right\rangle
	,\ \ \left\vert \Psi\left(  t_{\mathrm{in}}\right)  \right\rangle =\left\vert
	\Psi\right\rangle _{\mathrm{in}}\,.\label{a.1}%
\end{equation}
Here, $\hat{H}\left(  t\right)  $ is the Hamiltonian of the quantized
electromagnetic field $\hat{A}^{\mu}\left(  x\right)  =\left(  A^{0}\left(
x\right)  ,\mathbf{\hat{A}}\left(  \mathbf{r}\right)  \right)  $ interacting
with a classical four-current $j^{\mu}=\left(  j^{0}\left(  x\right)
,\mathbf{j}\left(  x\right)  \right)  $\footnote{We consider the Minkowski
	spacetime, $\eta_{\mu\nu}=\mathrm{diag}\left(  +1,-1,-1,-1\right)  ,$
	parameterized by coordinates $x^{\mu}=\left(  x^{0}=ct,\mathbf{r}\right)  $.
	Boldface letters denote three-dimensional vectors, e.g. $\mathbf{r}=\left(
	x^{i},\ i=1,2,3\right)  $, and Gaussian units are used.}. We can split the
potential $\mathbf{\hat{A}}\left(  \mathbf{r}\right)  $ into its transverse
$\mathbf{\hat{A}}_{\perp}\left(  \mathbf{r}\right)  $ and longitudinal parts
$\mathbf{\hat{A}}_{\parallel}\left(  \mathbf{r}\right)  $:%
\begin{align*}
	&  \mathbf{\hat{A}}\left(  \mathbf{r}\right)  =\mathbf{\hat{A}}_{\perp}\left(
	\mathbf{r}\right)  +\mathbf{\hat{A}}_{\parallel}\left(  \mathbf{r}\right)
	\ ,\\
	&  \mathbf{\hat{A}}_{\perp}\left(  \mathbf{r}\right)  =\delta_{\perp
	}\mathbf{\hat{A}}\left(  \mathbf{r}\right)  ,\ \ \mathbf{\hat{A}}_{\parallel
	}\left(  \mathbf{r}\right)  =(1-\delta_{\perp})\mathbf{\hat{A}}\left(
	\mathbf{r}\right)  \ ,\\
	&  \operatorname{div}\mathbf{\hat{A}}_{\perp}\left(  \mathbf{r}\right)
	=0,\ \ \operatorname{curl}\mathbf{\hat{A}}_{\parallel}\left(  \mathbf{r}%
	\right)  =0\ ,
\end{align*}
in which $\delta_{\perp}^{sp}=\delta^{sp}-\Delta^{-1}\partial^{s}\partial^{p}$
denotes the transverse projection operator,\ $(\delta_{\perp}\hat
{A}(\mathbf{r}))^{s}=\delta_{\perp}^{sp}A^{p}(\mathbf{r})$ (see Refs.
\cite{BogSh80,Greiner}).

Hereafter we work in the Coulomb gauge, in which the longitudinal degree of
freedom of $\mathbf{\hat{A}}\left(  \mathbf{r}\right)  $ is absent and only
transverse photons are present. Thus, $\mathbf{\hat{A}}\left(  \mathbf{r}%
\right)  =\mathbf{\hat{A}}_{\perp}\left(  \mathbf{r}\right)  $. The scalar
potential $A^{0}\left(  x\right)  $ is a non operatorial solution of the
Poisson equation,%
\begin{equation}
	\Delta A^{0}\left(  x\right)  =-\frac{4\pi}{c}j^{0}\left(  x\right)
	\rightarrow A^{0}\left(  x\right)  =\frac{1}{c}\int\frac{j^{0}\left(
		t,\mathbf{r}^{\prime}\right)  }{\left\vert \mathbf{r-r}^{\prime}\right\vert
	}d\mathbf{r}^{\prime}\,,\label{a.2}%
\end{equation}
while the operator of the vector potential $\mathbf{\hat{A}}\left(
\mathbf{r}\right)  $ reads:%
\begin{align}
	&  \mathbf{\hat{A}}\left(  \mathbf{r}\right)  =\sqrt{4\pi\hslash c}%
	\sum_{\lambda=1}^{2}\int\left[  \hat{c}_{\mathbf{k}\lambda}\mathbf{f}%
	_{\mathbf{k}\lambda}\left(  \mathbf{r}\right)  +\hat{c}_{\mathbf{k}\lambda
	}^{\dag}\mathbf{f}_{\mathbf{k}\lambda}^{\ast}\left(  \mathbf{r}\right)
	\right]  d\mathbf{k}\,,\ \ \operatorname{div}\mathbf{\hat{A}}\left(
	\mathbf{r}\right)  =0\,,\nonumber\\
	&  \mathbf{f}_{\mathbf{k}\lambda}\left(  \mathbf{r}\right)  =\frac
	{e^{i\mathbf{kr}}\boldsymbol{\epsilon}_{\mathbf{k}\lambda}}{\sqrt
		{2k_{0}\left(  2\pi\right)  ^{3}}}\,,\ \ k^{\mu}=\left(  k_{0}=\frac{\omega
	}{c},\mathbf{k}\right)  ,\ \ k_{0}=\left\vert \mathbf{k}\right\vert
	\,.\label{a.3}%
\end{align}
Here $\hat{c}_{\mathbf{k}\lambda}^{\dag}$ and $\hat{c}_{\mathbf{k}\lambda}$
are creation and annihilation operators of free photons with momenta
$\mathbf{k}$ and polarizations $\lambda$, satisfying the Bose-type commutation
relations%
\begin{equation}
	\left[  \hat{c}_{\mathbf{k}\lambda},\hat{c}_{\mathbf{k}^{\prime}%
		\lambda^{\prime}}^{\dag}\right]  =\delta_{\lambda\lambda^{\prime}}%
	\delta\left(  \mathbf{k}-\mathbf{k}^{\prime}\right)  ,\ \ \left[  \hat
	{c}_{\mathbf{k}\lambda},\hat{c}_{\mathbf{k}^{\prime}\lambda^{\prime}}\right]
	=\left[  \hat{c}_{\mathbf{k}\lambda}^{\dagger},\hat{c}_{\mathbf{k}^{\prime
		}\lambda^{\prime}}^{\dag}\right]  =0\,.\label{a.4}%
\end{equation}
The complex polarization vectors $\boldsymbol{\epsilon}_{\mathbf{k}\lambda}$
satisfy the orthogonality and completeness relations:%
\begin{equation}
	\boldsymbol{\epsilon}_{\mathbf{k}\lambda}\boldsymbol{\epsilon}_{\mathbf{k}%
		\lambda^{\prime}}^{\ast}=\delta_{\lambda\lambda^{\prime}}%
	,\ \ \boldsymbol{\epsilon}_{\mathbf{k}\lambda}\mathbf{k}=0,\ \ \sum
	_{\lambda=1}^{2}\epsilon_{\mathbf{k}\lambda}^{i}\epsilon_{\mathbf{k}\lambda
	}^{j\ast}=\delta^{ij}-n^{i}n^{j},\ \ n^{i}=\frac{k^{i}}{\left\vert
		\mathbf{k}\right\vert }\,.\label{a.4b}%
\end{equation}
The Hamiltonian $\hat{H}\left(  t\right)  $ reads:%
\begin{equation}
	\hat{H}\left(  t\right)  =\hat{H}_{\gamma}+\frac{1}{c}\int\left[  \frac{1}%
	{2}j_{0}\left(  x\right)  A^{0}\left(  x\right)  -\mathbf{j}\left(  x\right)
	\mathbf{\hat{A}}\left(  \mathbf{r}\right)  \right]  d\mathbf{r},\ \ \hat
	{H}_{\gamma}=\hslash c\sum_{\lambda=1}^{2}\int k_{0}\hat{c}_{\mathbf{k}%
		\lambda}^{\dag}\hat{c}_{\mathbf{k}\lambda}d\mathbf{k\ }.\label{a.5}%
\end{equation}

One can demonstrate that a solution of equation (\ref{a.1}) can be written as%
\begin{equation}
	\left\vert \Psi\left(  t\right)  \right\rangle =U\left(  t,t_{\mathrm{in}%
	}\right)  \left\vert \Psi\right\rangle _{\mathrm{in}}\ ,\label{a.5a}%
\end{equation}
where the evolution operator $U\left(  t,t_{\mathrm{in}}\right)  $ has the
form (see e.g. \cite{BagGiSF20}):%
\begin{align}
	&  U\left(  t,t_{\mathrm{in}}\right)  =U_{\gamma}\left(  t,t_{\mathrm{in}%
	}\right)  \exp\left[  -\frac{i}{\hslash}\hat{B}\left(  t\right)  \right]
	,\ \ U_{\gamma}\left(  t,t_{\mathrm{in}}\right)  =\exp\left[  -\frac
	{i}{\hslash}\hat{H}_{\gamma}\left(  t-t_{\mathrm{in}}\right)  \right]
	\,,\nonumber\\
	&  \hat{B}\left(  t\right)  =\frac{1}{c}\int_{0}^{t}dt^{\prime}\int\left\{
	\frac{1}{2}j_{0}\left(  x^{\prime}\right)  A^{0}\left(  x^{\prime}\right)
	-\mathbf{j}\left(  x^{\prime}\right)  \left[  \mathbf{\hat{A}}\left(
	x^{\prime}\right)  +\frac{1}{2}\mathbf{\tilde{A}}\left(  x^{\prime}\right)
	\right]  \right\}  d\mathbf{r}^{\prime}\ ,\nonumber\\
	&  \mathbf{\tilde{A}}\left(  x\right)  =4\pi\int_{t_{\mathrm{in}}}%
	^{t}dt^{\prime}\int D_{0}\left(  x-x^{\prime}\right)  \mathbf{j}_{\perp
	}\left(  x^{\prime}\right)  d\mathbf{r}^{\prime},\ \ j_{\perp}^{s}%
	(x)=\delta_{\perp}^{sp}j^{p}(x)\,,\ \ \operatorname{div}\mathbf{j}_{\perp
	}(x)=0\ ,\nonumber\\
	&  \mathbf{\hat{A}}\left(  x\right)  =U_{\gamma}^{\dagger}\left(
	t,t_{\mathrm{in}}\right)  \mathbf{\hat{A}}\left(  \mathbf{r}\right)
	U_{\gamma}\left(  t,t_{\mathrm{in}}\right)  =\sqrt{4\pi c\hbar}\sum
	\limits_{\lambda=1}^{2}\int\left[  \hat{c}_{\mathbf{k}\lambda}\mathbf{f}%
	_{\mathbf{k}\lambda}\left(  x,t_{\mathrm{in}}\right)  +\hat{c}_{\mathbf{k}%
		\lambda}^{\dag}\mathbf{f}_{\mathbf{k}\lambda}^{\ast}\left(  x,t_{\mathrm{in}%
	}\right)  \right]  d\mathbf{k}\ ,\nonumber\\
	&  \ \mathbf{f}_{\mathbf{k}\lambda}\left(  x,t_{\mathrm{in}}\right)
	=\mathbf{f}_{\mathbf{k}\lambda}\left(  \mathbf{r}\right)  \exp\left[
	-ik_{0}c(t-t_{\mathrm{in}})\right]  \ .\label{a.6}%
\end{align}
Here, $D_{0}\left(  x-x^{\prime}\right)  $ is the Pauli-Jordan singular
function with zero mass (see, e.g. Refs. \cite{Greiner, BogSh80}),%
\begin{align}
	&  D_{0}\left(  x-x^{\prime}\right)  =\frac{1}{\left(  2\pi\right)  ^{3}}%
	\int\frac{\sin k(x-x^{\prime})}{k_{0}}d\mathbf{k}\,,\nonumber\\
	&  [\hat{A}^{s}(x),\hat{A}^{p}(x^{\prime})]=-4i\pi\hslash c\ \delta_{\perp
	}^{sp}\ D_{0}\left(  x-x^{\prime}\right)  \ .\label{1.1}%
\end{align}

Finally, we point out that the evolution operator $U\left(  t,t_{\mathrm{in}%
}\right)  $ can be alternatively presented in the following form:%
\begin{align}
	&  U\left(  t,t_{\mathrm{in}}\right)  =\exp\left[  i\phi\left(
	t,t_{\mathrm{in}}\right)  \right]  U_{\gamma}\left(  t,t_{\mathrm{in}}\right)
	\mathcal{D}\left(  y\right)  \,,\nonumber\\
	&  \phi\left(  t,t_{\mathrm{in}}\right)  =-\frac{1}{2\hslash c}\int%
	_{t_{\mathrm{in}}}^{t}dt^{\prime}\int\left[  j_{0}\left(  x^{\prime}\right)
	A^{0}\left(  x^{\prime}\right)  -\mathbf{j}\left(  x^{\prime}\right)
	\mathbf{\tilde{A}}\left(  x^{\prime}\right)  \right]  d\mathbf{r}^{\prime
	}\,,\nonumber\\
	&  \mathcal{D}\left(  y\right)  =\exp\sum_{\lambda=1}^{2}\int\left[
	y_{\mathbf{k}\lambda}\left(  t,t_{\mathrm{in}}\right)  \hat{c}_{\mathbf{k}%
		\lambda}^{\dag}-y_{\mathbf{k}\lambda}^{\ast}\left(  t,t_{\mathrm{in}}\right)
	\hat{c}_{\mathbf{k}\lambda}\right]  d\mathbf{k}\,,\nonumber\\
	&  y_{\mathbf{k}\lambda}\left(  t,t_{\mathrm{in}}\right)  =i\sqrt{\frac{4\pi
		}{\hslash c}}\int_{t_{\mathrm{in}}}^{t}dt^{\prime}\int\mathbf{j}\left(
	x^{\prime}\right)  \mathbf{f}_{\mathbf{k}\lambda}^{\ast}\left(  x^{\prime
	},t_{\mathrm{in}}\right)  d\mathbf{r}^{\prime}\,. \label{a.6a}%
\end{align}

Let us calculate the mean value of the electromagnetic field operator
$\mathbf{\hat{A}}\left(  \mathbf{r}\right)  $ with respect to state
(\ref{a.5a}). To this end, we use Eqs. (\ref{a.6}), (\ref{a.6a}) and the
identities%
\begin{equation}
	\mathcal{D}^{\dagger}\left(  y\right)  \hat{c}_{\mathbf{k}\lambda}%
	\mathcal{D}\left(  y\right)  =y_{\mathbf{k}\lambda}\left(  t,t_{\mathrm{in}%
	}\right)  +\hat{c}_{\mathbf{k}\lambda},\ \ U_{\gamma}^{\dagger}\left(
	t_{\mathrm{in}},t\right)  \hat{c}_{\mathbf{k}\lambda}U_{\gamma}\left(
	t_{\mathrm{in}},t\right)  =\hat{c}_{\mathbf{k}\lambda}e^{-ik_{0}c\left(
		t-t_{\mathrm{in}}\right)  }\,.\label{a.7a}%
\end{equation}
With their help, we find:%
\begin{align}
	&  \left\langle \Psi\left(  t\right)  \left\vert \mathbf{\hat{A}}\left(
	\mathbf{r}\right)  \right\vert \Psi\left(  t\right)  \right\rangle
	=\ _{\mathrm{in}}\left\langle \Psi\left\vert \mathcal{D}^{\dagger}\left(
	y\right)  U_{\gamma}^{\dagger}\left(  t_{\mathrm{in}},t\right)  \mathbf{\hat
		{A}}\left(  \mathbf{r}\right)  U_{\gamma}\left(  t_{\mathrm{in}},t\right)
	\mathcal{D}\left(  y\right)  \right\vert \Psi\right\rangle _{\mathrm{in}%
	}\nonumber\\
	&  =\ _{\mathrm{in}}\left\langle \Psi\left\vert \mathcal{D}^{\dagger}\left(
	y\right)  \mathbf{\hat{A}}\left(  x\right)  \mathcal{D}\left(  y\right)
	\right\vert \Psi\right\rangle _{\mathrm{in}}\nonumber\\
	&  =\mathbf{A}_{\mathrm{free}}\left(  x\right)  +2\sqrt{4\pi\hslash
		c}\operatorname{Re}\sum_{\lambda=1}^{2}\int y_{\mathbf{k}\lambda}\left(
	t,t_{\mathrm{in}}\right)  \mathbf{f}_{\mathbf{k}\lambda}\left(
	x,t_{\mathrm{in}}\right)  d\mathbf{k}\,.\label{a.7d}%
\end{align}
The first term $\mathbf{A}_{\mathrm{free}}\left(  x\right)  =\ _{\mathrm{in}%
}\left\langle \Psi\left\vert \mathbf{\hat{A}}\left(  x\right)  \right\vert
\Psi\right\rangle _{\mathrm{in}}$ is a free electromagnetic field satisfying
the initial condition $\mathbf{A}_{\mathrm{free}}\left(  \mathbf{r,}%
t_{\mathrm{in}}\right)  =\ _{\mathrm{in}}\left\langle \Psi\left\vert
\mathbf{\hat{A}}\left(  \mathbf{r}\right)  \right\vert \Psi\right\rangle
_{\mathrm{in}}$. If the initial state $\left\vert \Psi\right\rangle
_{\mathrm{in}}$ is a state with a definite number of the photons, then
$\mathbf{A}_{\mathrm{in}}\left(  x\right)  $ is zero.

As for the second term in the last line of Eq. (\ref{a.7d}), we, using Eqs.
(\ref{a.6}) and (\ref{a.4b}), may write:%
\begin{equation}
	\sum_{\lambda=1}^{2}\int y_{\mathbf{k}\lambda}\left(  t,t_{\mathrm{in}%
	}\right)  f_{\mathbf{k}\lambda}^{s}\left(  x,t_{\mathrm{in}}\right)
	d\mathbf{k}=\frac{i}{2}\sqrt{\frac{4\pi}{\hslash c}}\int\frac{d\mathbf{k}%
	}{\left(  2\pi\right)  ^{3}}\frac{1}{k_{0}}\int_{t_{\mathrm{in}}}%
	^{t}dt^{\prime}\int j^{p}\left(  x^{\prime}\right)  \left[  \delta_{\perp
	}^{ps}e^{ik\left(  x^{\prime}-x\right)  }\right]  d\mathbf{r}^{\prime
	}\,.\label{w1}%
\end{equation}
Perfoming the integration by parts over $\mathbf{r}^{\prime}$ in the RHS of
Eq. (\ref{w1}), assuming that external currents vanish at remote boundaries,
and recalling definition (\ref{a.6}), we find:%
\begin{align}
	&  2\sqrt{4\pi\hslash c}\operatorname{Re}\sum_{\lambda=1}^{2}\int
	y_{\mathbf{k}\lambda}\left(  t,t_{\mathrm{in}}\right)  \mathbf{f}%
	_{\mathbf{k}\lambda}\left(  x,t_{\mathrm{in}}\right)  d\mathbf{k}=4\pi
	\int_{t_{\mathrm{in}}}^{t}dt^{\prime}\int d\mathbf{r}^{\prime}\mathbf{j}%
	_{\perp}\left(  x^{\prime}\right)  \int\frac{\sin k\left(  x-x^{\prime
		}\right)  }{\left(  2\pi\right)  ^{3}k_{0}}d\mathbf{k}\nonumber\\
	&  =4\pi\int_{t_{\mathrm{in}}}^{t}dt^{\prime}\int D_{0}\left(  x-x^{\prime
	}\right)  \mathbf{j}_{\perp}\left(  x^{\prime}\right)  d\mathbf{r}^{\prime
	}=\mathbf{\tilde{A}}\left(  x\right)  \ ,\label{w3}%
\end{align}
where the field $\mathbf{\tilde{A}}\left(  x\right)  $ is given by Eq.
(\ref{a.6}). To recognize $\mathbf{\tilde{A}}\left(  x\right)  $ as the
Li?nard-Wiechert retarded vector potential $\mathbf{A}_{\mathrm{LW}}\left(
x\right)  $ in the Coulomb gauge,%
\begin{equation}
	\mathbf{A}_{\mathrm{LW}}\left(  x\right)  =\frac{1}{c}\int\frac{\mathbf{j}%
		_{\perp}\left(  t_{\mathrm{ret}},\mathbf{r}^{\prime}\right)  }{\left\vert
		\mathbf{r}-\mathbf{r}^{\prime}\right\vert }d\mathbf{r}^{\prime}%
	,\ \ t_{\mathrm{ret}}=t-\frac{\left\vert \mathbf{r-r}^{\prime}\right\vert }%
	{c}\,,\label{w4}%
\end{equation}
see e.g. Ref. \cite{LanLi71}, it is enough to use the following representation
of the Pauli-Jordan singular function (see Refs. \cite{Greiner, BogSh80})%
\begin{equation}
	D_{0}\left(  x-x^{\prime}\right)  =\frac{\delta\left(  \left\vert
		\mathbf{r}-\mathbf{r}^{\prime}\right\vert -c(t-t^{\prime})\right)
		-\delta\left(  \left\vert \mathbf{r}-\mathbf{r}^{\prime}\right\vert
		+c(t-t^{\prime})\right)  }{4\pi\left\vert \mathbf{r}-\mathbf{r}^{\prime
		}\right\vert }\,,\label{1.2}%
\end{equation}
in Eq. (\ref{a.6}) for $\mathbf{\tilde{A}}\left(  x\right)  $,%
\begin{eqnarray}
	\mathbf{\tilde{A}}\left(  x\right)  &=&\frac{1}{c}\int_{t_{\mathrm{in}}%
	}^{t}dt^{\prime}\int\delta\left(  t^{\prime}-t+\frac{\left\vert \mathbf{r}%
		-\mathbf{r}^{\prime}\right\vert }{c}\right)  \frac{\mathbf{j}_{\perp}\left(
		t^{\prime},\mathbf{r}^{\prime}\right)  }{\left\vert \mathbf{r}-\mathbf{r}%
		^{\prime}\right\vert }d\mathbf{r}^{\prime}\nonumber\\
	  &=&\frac{1}{c}\int\frac{\mathbf{j}_{\perp}\left(  t_{\mathrm{ret}}%
		,\mathbf{r}^{\prime}\right)  }{\left\vert \mathbf{r}-\mathbf{r}^{\prime
		}\right\vert }d\mathbf{r}^{\prime}=\mathbf{A}_{\mathrm{LW}}\left(  x\right)
	\ ,\label{a.7c}%
\end{eqnarray}
to conclude that%
\begin{equation}
	\left\langle \Psi\left(  t\right)  \left\vert \mathbf{\hat{A}}\left(
	\mathbf{r}\right)  \right\vert \Psi\left(  t\right)  \right\rangle
	=\mathbf{A}_{\mathrm{free}}\left(  x\right)  +\mathbf{A}_{\mathrm{LW}}\left(
	x\right)  \,.\label{1.3}%
\end{equation}
It should be noted that the absence of the longitudinal component of the
current density $\mathbf{j}_{\perp}\left(  t_{\mathrm{ret}},\mathbf{r}%
^{\prime}\right)  $ the potential $\mathbf{A}_{\mathrm{LW}}\left(  x\right)  $
is transversal, $\operatorname{div}\mathbf{A}_{\mathrm{LW}}\left(  x\right)
=0$ and satisfies the Maxwell equation%
\begin{equation}
	\square\mathbf{A}_{\mathrm{LW}}\left(  x\right)  =\frac{4\pi}{c}%
	\mathbf{j}_{\perp}\left(  x\right)  \,,\label{a.7.6}%
\end{equation}
as it should be in the Coulomb gauge under consideration.

\subsection{Photon emission characteristics\label{S2.2}}

Having in hands the state vector $\left\vert \Psi\left(  t\right)
\right\rangle $, we can calculate probability amplitudes of photon emission
from different initial states. For example, assuming that the initial state is
the vacuum state, $\left\vert \Psi\left(  t_{\mathrm{in}}\right)
\right\rangle =\left\vert 0\right\rangle $, so that%
\begin{equation}
	\left\vert \Psi\left(  t\right)  \right\rangle =\exp\left[  i\phi\left(
	t,t_{\mathrm{in}}\right)  \right]  U_{\gamma}\left(  t,t_{\mathrm{in}}\right)
	\mathcal{D}\left(  y\right)  \left\vert 0\right\rangle \,,\label{2.1}%
\end{equation}
the transition probability amplitude $R\left(  \left\{  N\right\}
;t,t_{\mathrm{in}}\right)  $ to the state%
\begin{equation}
	\left\vert \left\{  N\right\}  \right\rangle =\frac{1}{\sqrt{N!}}\prod
	_{a=1}^{N}\hat{c}_{\mathbf{k}_{a}\lambda_{a}}^{\dag}\left\vert 0\right\rangle
	\,,\label{2.2}%
\end{equation}
with $N$ photons with momenta $\mathbf{k}_{a}$ and polarizations $\lambda_{a}%
$, $a=1,2...,N$, has the form:%
\begin{eqnarray}
	  R\left(  \left\{  N\right\}  ;t,t_{\mathrm{in}}\right)  &=&\left\langle
	\left\{  N\right\}  |\Psi\left(  t\right)  \right\rangle =\frac{R\left(
		0;t,t_{\mathrm{in}}\right)  }{\sqrt{N!}}\prod_{a=1}^{N}y_{\mathbf{k}%
		_{a}\lambda_{a}}\left(  t,t_{\mathrm{in}}\right)  \exp\left[  -i\left\vert
	\mathbf{k}_{a}\right\vert c\left(  t-t_{\mathrm{in}}\right)  \right]
	\,,\nonumber\\
	  R\left(  0;t,t_{\mathrm{in}}\right)  &=&\left\langle 0|\Psi\left(  t\right)
	\right\rangle =\exp\left[  i\phi\left(  t,t_{\mathrm{in}}\right)  -\frac{1}%
	{2}\sum_{\lambda=1}^{2}\int\left\vert y_{\mathbf{k}\lambda}\left(
	t,t_{\mathrm{in}}\right)  \right\vert ^{2}d\mathbf{k}\right]  \,.\label{a.8}%
\end{eqnarray}
Here $R\left(  0;t,t_{\mathrm{in}}\right)  $ is the transition probability
amplitude with zero photon emission. The corresponding transition probability
reads:%
\begin{eqnarray}
	 P\left(  \left\{  N\right\}  ;t,t_{\mathrm{in}}\right)  &=&\left\vert
	R\left(  \left\{  N\right\}  ;t,t_{\mathrm{in}}\right)  \right\vert
	^{2}=P\left(  0;t,t_{\mathrm{in}}\right)  p\left(  \left\{  N\right\}
	;t,t_{\mathrm{in}}\right)  \,,\nonumber\\
	P\left(  0;t,t_{\mathrm{in}}\right)  &=&\exp\left[  -\sum_{\lambda=1}^{2}%
	\int\left\vert y_{\mathbf{k}\lambda}\left(  t,t_{\mathrm{in}}\right)
	\right\vert ^{2}d\mathbf{k}\right]  \,,\nonumber\\
	p\left(  \left\{  N\right\}
	;t,t_{\mathrm{in}}\right)  &=&\frac{1}{N!}\prod_{a=1}^{N}\left\vert
	y_{\mathbf{k}_{a}\lambda_{a}}\left(  t,t_{\mathrm{in}}\right)  \right\vert
	^{2}\,.\label{a.9}%
\end{eqnarray}

With the aid of the above probability and bearind in mind that the
electromagnetic energy of $N$ photons is%
\begin{equation}
	W\left(  \left\{  N\right\}  \right)  =\hslash c\sum_{a=1}^{N}k_{0,a}%
	\,,\ \ k_{0,a}=\left\vert \mathbf{k}_{a}\right\vert \,,\label{f8}%
\end{equation}
the total electromagnetic energy of $N$ photons\footnote{that is, the
	electromagnetic energy of $N$ photons with all possible momenta and
	polarizations.} radiated by the source, $W\left(  N;t,t_{\mathrm{in}}\right)
$, is the sum of differential energies (\ref{f8}) weighted by the probability
$P\left(  \left\{  N\right\}  ;t,t_{\mathrm{in}}\right)  $,%
\begin{eqnarray}
	W\left(  N;t,t_{\mathrm{in}}\right)  &=&\sum_{\left\{  N\right\}  }W\left(
	\left\{  N\right\}  \right)  P\left(  \left\{  N\right\}  ;t,t_{\mathrm{in}%
	}\right)  \nonumber\\
    &=&\frac{W\left(  1;t,t_{\mathrm{in}}\right)  }{\left(  N-1\right)
		!}\left[  \sum_{\lambda=1}^{2}\int\left\vert y_{\mathbf{k}\lambda}\left(
	t,t_{\mathrm{in}}\right)  \right\vert ^{2}d\mathbf{k}\right]  ^{N-1}%
	\,.\label{f9}%
\end{eqnarray}
Here, the sum indexed with $\left\{  N\right\}  $ represents a summation over
quantum numbers of $N$ photons,%
\begin{equation}
	\sum_{\left\{  N\right\}  }=\prod_{a=1}^{N}\left(  \sum_{\lambda_{a}=1}%
	^{2}\int d\mathbf{k}_{a}\right)  ,\ \ \sum_{N=1}^{\infty}\sum_{\left\{
		N\right\}  }P\left(  \left\{  N\right\}  ;t,t_{\mathrm{in}}\right)
	=1\,,\label{2.3}%
\end{equation}
while $W\left(  1;t,t_{\mathrm{in}}\right)  $ denotes the electromagnetic
energy of one photon radiated by the external current,%
\begin{equation}
	W\left(  1;t,t_{\mathrm{in}}\right)  =\hslash c\ P\left(  0;t,t_{\mathrm{in}%
	}\right)  \sum_{\lambda=1}^{2}\int k_{0}\left\vert y_{\mathbf{k}\lambda
	}\left(  t,t_{\mathrm{in}}\right)  \right\vert ^{2}d\mathbf{k}\,.\label{f10}%
\end{equation}
Finally, summing Eq. (\ref{f9}) over all the photons, we obtain the total
electromagnetic energy radiated by the external current:%
\begin{eqnarray}
	W\left(  t,t_{\mathrm{in}}\right)  &=&\sum_{N=1}^{\infty}W\left(
	N;t,t_{\mathrm{in}}\right)  \nonumber\\
	&=&W\left(  1;t,t_{\mathrm{in}}\right)  \exp\left[
	\sum_{\lambda=1}^{2}\int\left\vert y_{\mathbf{k}\lambda}\left(
	t,t_{\mathrm{in}}\right)  \right\vert ^{2}d\mathbf{k}\right]\nonumber\\
	  &=&\hslash
	c\sum_{\lambda=1}^{2}\int k_{0}\left\vert y_{\mathbf{k}\lambda}\left(
	t,t_{\mathrm{in}}\right)  \right\vert ^{2}d\mathbf{k}\,.\label{f10b}%
\end{eqnarray}

With these results we may define the rate in which the energy is emitted from
the souce by differentiating the energy (\ref{f10b}) with respect to time,%
\begin{equation}
	w\left(  t,t_{\mathrm{in}}\right)  =\frac{\partial W\left(  t,t_{\mathrm{in}%
		}\right)  }{\partial t}=2\hslash c\sum_{\lambda=1}^{2}\int k_{0}%
	\operatorname{Re}\left[  y_{\mathbf{k}\lambda}\left(  t,t_{\mathrm{in}%
	}\right)  \frac{\partial}{\partial t}y_{\mathbf{k}\lambda}^{\ast}\left(
	t,t_{\mathrm{in}}\right)  \right]  d\mathbf{k}\,. \label{f14}%
\end{equation}

To facilitate the comparison with results from classical electrodynamics, it
is convenient to express the total energy (\ref{f10}) in terms of the external
current $\mathbf{j}\left(  x\right)  $. To this end, we compute the modulus
square of $y_{\mathbf{k}\lambda}\left(  t,t_{\mathrm{in}}\right)  $ and sum
the result over all photon polarizations to discover that the electromagnetic
energies (\ref{f10}), (\ref{f10b}) admit the forms%
\begin{eqnarray}
	  W\left(  t,t_{\mathrm{in}}\right)  &=&4\pi^{2}\int\left\vert \mathbf{n}%
	\times\left[  \mathbf{n}\times\mathbf{\tilde{j}}\left(  k\mathbf{;}%
	t,t_{\mathrm{in}}\right)  \right]  \right\vert ^{2}d\mathbf{k}\,,\nonumber\\
	  W\left(  1;t,t_{\mathrm{in}}\right)  &=&W\left(  t,t_{\mathrm{in}}\right)
	\exp\left[  -\frac{1}{\hslash c}W\left(  t,t_{\mathrm{in}}\right)  \right]
	\ ,\nonumber\\
	  \mathbf{\tilde{j}}\left(  k\mathbf{;}t,t_{\mathrm{in}}\right)  &=&\frac
	{1}{\sqrt{2\pi}}\int_{t_{\mathrm{in}}}^{t}e^{ik_{0}ct^{\prime}}\mathbf{\tilde
		{j}}\left(  \mathbf{k};t^{\prime}\right)  dt^{\prime}\ , \label{f12}%
\end{eqnarray}
where $\mathbf{\tilde{j}}\left(  \mathbf{k};t^{\prime}\right)  $ is the
three-dimensional Fourier transform of the current density,%
\begin{equation}
	\mathbf{\tilde{j}}\left(  \mathbf{k};t^{\prime}\right)  =\frac{1}{(2\pi
		)^{3/2}}\int e^{-i\mathbf{kr}^{\prime}}\mathbf{j}\left(  x^{\prime}\right)
	d\mathbf{r}^{\prime}\,. \label{f13}%
\end{equation}
From these results we may conclude that Eq. (\ref{f10b}) coincides with the
expression for the classical electromagnetic energy in the limits
$t_{\mathrm{in}}\rightarrow-\infty$, $t\rightarrow+\infty$, namely,%
\begin{align}
	&  W\left(  +\infty,-\infty\right)  =W_{\mathrm{cl}}\,,\label{f14a}\\
	&  W\left(  1;+\infty,-\infty\right)  =W_{\mathrm{cl}}\exp\left[  -\frac
	{1}{\hslash c}W_{\mathrm{cl}}\right]  \,,\label{f14b}\\
	&  W_{\mathrm{cl}}=4\pi^{2}\int\left\vert \mathbf{n}\times\left[
	\mathbf{n}\times\mathbf{\tilde{j}}\left(  k\right)  \right]  \right\vert
	^{2}d\mathbf{k}\,,\ \ \mathbf{\tilde{j}}\left(  k\right)  =\frac{1}{\sqrt
		{2\pi}}\int_{-\infty}^{+\infty}e^{ik_{0}ct^{\prime}}\mathbf{\tilde{j}}\left(
	\mathbf{k};t^{\prime}\right)  dt^{\prime}\,. \label{f14c}%
\end{align}
Here, $W_{\mathrm{cl}}$ denotes precisely the spectral decomposition of the
electromagnetic energy emitted by the charge in the classical theory of
radiation \cite{LanLi71,Jacks99}.

Last, but not least, we may sum Eq. (\ref{f14}) over the photon polarizations
(or, equivalently, differentiate Eq. (\ref{f12}) with respect to time) to
express the energy rate in terms of the external current as follows:%
\begin{equation}
	w\left(  t,t_{\mathrm{in}}\right)  =2\left(  2\pi\right)  ^{3/2}%
	\int\operatorname{Re}e^{-ik_{0}ct}\left\{  \mathbf{\tilde{j}}^{\ast}\left(
	\mathbf{k};t\right)  \mathbf{\tilde{j}}\left(  k\mathbf{;}t,t_{\mathrm{in}%
	}\right)  -\left[  \mathbf{n\tilde{j}}^{\ast}\left(  \mathbf{k};t\right)
	\right]  \left[  \mathbf{n\tilde{j}}\left(  k\mathbf{;}t,t_{\mathrm{in}%
	}\right)  \right]  \right\}  d\mathbf{k}\,.\label{f15}%
\end{equation}
In contrast to the energy (\ref{f12}), the rate (\ref{f15}) is asymmetrical
with respect to time because the time derivative in (\ref{f14}) affects only
the upper integration limit of function $y_{\mathbf{k}\lambda}\left(
t,t_{\mathrm{in}}\right)  $. As a result, the energy rates computed from
(\ref{f14}) or (\ref{f15}) shall feature the same type of asymmetry. However,
it is possible to define a symmetrical energy rate choosing
$t=T/2=-t_{\mathrm{in}}$ and differentiating the corresponding energy with
respect to $T$: $w\left(  T\right)  \equiv\frac{\partial}{\partial T}W\left(
+T/2,-T/2\right)  $. Following this convention, the energy rate (\ref{f15})
admits the form%
{\footnotesize
\begin{eqnarray}
	w\left(  T\right)  &=&(2\pi)^{3/2}\operatorname{Re}\int e^{ik_{0}cT/2}\left[
	\mathbf{\tilde{j}}\left(  \mathbf{k},+T/2\right)  \mathbf{\tilde{j}}^{\ast
	}\left(  k\mathbf{;}T\right)  +\mathbf{\tilde{j}}^{\ast}\left(  \mathbf{k}%
	,-T/2\right)  \mathbf{\tilde{j}}\left(  k\mathbf{;}T\right)  \right]
	d\mathbf{k}\nonumber\\
	&-& (2\pi)^{3/2}\operatorname{Re}\int e^{ik_{0}cT/2}\left\{  \left[
	\mathbf{n\tilde{j}}\left(  \mathbf{k},+T/2\right)  \right]  \left[
	\mathbf{n\tilde{j}}^{\ast}\left(  k\mathbf{;}T\right)  \right]  +\left[
	\mathbf{n\tilde{j}}^{\ast}\left(  \mathbf{k},-T/2\right)  \right]  \left[
	\mathbf{n\tilde{j}}\left(  k\mathbf{;}T\right)  \right]  \right\}
	d\mathbf{k\,,}\nonumber\\
	  \mathbf{\tilde{j}}\left(  k\mathbf{;}T\right)  &=&\frac{1}{\sqrt{2\pi}}%
	\int_{-T/2}^{T/2}e^{ik_{0}ct^{\prime}}\mathbf{\tilde{j}}\left(  \mathbf{k}%
	;t^{\prime}\right)  dt^{\prime},\ \ \mathbf{\tilde{j}}\left(  \mathbf{k};\pm
	T/2\right)  =\frac{1}{(2\pi)^{3/2}}\int e^{-i\mathbf{kr}^{\prime}}%
	\mathbf{j}\left(  \pm T/2,\mathbf{r}^{\prime}\right)  d\mathbf{r}^{\prime
	}\,.\label{f16}%
\end{eqnarray}
}

\section{Electromagnetic energies and rates radiated by an accelerated
	charge\label{S3}}

\subsection{Total electromagnetic energy\label{Sec3.1}}

The electromagnetic energies and rates discussed so far apply to any charge
distribution and the force field accelerating them. However, for concrete
computations, it is convenient specifying both the external force and the
current density. Thus, from now on we focus on studying the electromagnetic
radiation emitted by a pointlike charged particle (with algebraic charge $q$
and mass $m$) accelerated by a constant and homogeneous electric field,
$\mathbf{E}\left(  x\right)  =\mathbf{E}=\mathrm{const.}$ in the framework of
the semiclassical theory. The current density describing such a particle
moving with speed $\mathbf{v}\left(  t\right)  =d\mathbf{r}\left(  t\right)
/dt$ is $\mathbf{j}\left(  x\right)  =qc\boldsymbol{\beta}\left(  t\right)
\delta\left(  \mathbf{r}-\mathbf{r}\left(  t\right)  \right)  $,
$\boldsymbol{\beta}\left(  t\right)  =\mathbf{v}\left(  t\right)  /c$.
Plugging this current into Eq. (\ref{a.6a}) we find,%
\begin{equation}
	y_{\mathbf{k}\lambda}\left(  t,t_{\mathrm{in}}\right)  =-\frac{iqc}{2\pi}%
	\frac{e^{-ik_{0}ct_{\mathrm{in}}}}{\sqrt{\hslash ck_{0}}}\int_{t_{\mathrm{in}%
	}}^{t}e^{i\Phi\left(  t^{\prime}\right)  }\boldsymbol{\beta}\left(  t^{\prime
	}\right)  \boldsymbol{\epsilon}_{\mathbf{k}\lambda}^{\ast}dt^{\prime}%
	,\ \ \Phi\left(  t\right)  =k_{0}ct-\mathbf{kr}\left(  t\right)
	\,.\label{s3.1}%
\end{equation}
Without loss of generality, we assume that the electric field points to the
positive direction of the $z$-axis, $\mathbf{E}=\left(  0,0,E\right)  $. The
solutions to Lorentz equations for the particle in the external field under
consideration with initial position $\underline{\mathbf{r}}=\mathbf{r}\left(
0\right)  =\left(  \underline{x},\underline{y},\underline{z}\right)  $\ and
velocity $\underline{\mathbf{v}}=\mathbf{v}\left(  0\right)  =\left(
\underline{v}_{x},\underline{v}_{y},\underline{v}_{z}\right)  $ read:%
\begin{align}
	&  \mathbf{r}_{\perp}\left(  t\right)  =\mathbf{\tilde{r}}_{\perp}%
	+\frac{\underline{\mathbf{u}}_{\perp}}{\varepsilon}\mathrm{arc\sinh}\left(
	\frac{\varepsilon t+\underline{u}_{\Vert}/c}{\varrho}\right)
	,\ \ \boldsymbol{\beta}_{\perp}\left(  t\right)  =\frac{\underline{\mathbf{u}%
		}_{\perp}/c}{\sqrt{\varrho^{2}+\left(  \varepsilon t+\underline{u}_{\parallel
			}/c\right)  ^{2}}}\,,\nonumber\\
	&  r_{\parallel}\left(  t\right)  =\underline{r}_{\parallel}+\frac
	{c}{\varepsilon}\sqrt{\varrho^{2}+\left(  \varepsilon t+\underline{u}_{\Vert
		}/c\right)  ^{2}},\ \ \beta_{\parallel}\left(  t\right)  =\frac{\varepsilon
		t+\underline{u}_{\parallel}/c}{\sqrt{\varrho^{2}+\left(  \varepsilon
			t+\underline{u}_{\parallel}/c\right)  ^{2}}}\,,\nonumber\\
	&  \varepsilon=\frac{qE}{mc},\ \ \varrho=\frac{\sqrt{m^{2}c^{2}%
			+\underline{\mathbf{P}}_{\perp}^{2}}}{mc},\ \ \underline{\mathbf{P}}_{\perp
	}=m\underline{\mathbf{u}}_{\perp},\ \ \underline{\mathbf{u}}=\frac
	{c\underline{\boldsymbol{\beta}}}{\sqrt{1-\underline{\boldsymbol{\beta}}^{2}}%
	}\,.\label{s3.1b}%
\end{align}
Here, $\mathbf{\tilde{r}}_{\perp}=\underline{\mathbf{r}}_{\perp}-\left(
\underline{\mathbf{u}}_{\perp}/\varepsilon\right)  \mathrm{arc\sinh}\left(
\underline{u}_{\Vert}/\varrho c\right)  $ and the indexes \textquotedblleft%
$\perp$\textquotedblright, \textquotedblleft$\parallel$\textquotedblright%
\ label components \textquotedblleft perpendicular\textquotedblright,
\textquotedblleft parallel\textquotedblright\ to the external
field\footnote{e.g., $\mathbf{k}=\left(  \mathbf{k}_{\perp},k_{\parallel
	}\right)  $, $\mathbf{k}_{\perp}=\left(  k_{x},k_{y},0\right)  $,
	$k_{\parallel}=k_{z}$. The meanings of the symbols \textquotedblleft$\perp
	$\textquotedblright\ and \textquotedblleft$\parallel$\textquotedblright%
	\ employed here should not be confused with those used in subsection
	\ref{S2.1}.}, respectively. Using the above solutions, we substitute
$t^{\prime}$ by $\eta^{\prime}$%
\begin{equation}
	t^{\prime}=-\frac{\underline{u}_{\Vert}}{\varepsilon c}+\frac{\varrho
	}{\varepsilon}\sinh\eta^{\prime}\,,\label{s3.2}%
\end{equation}
and introduce three auxiliary variables\footnote{$z$ and $\xi$ can also be
	defined though the identities $z\sinh\xi=\left(  \varrho c/\varepsilon\right)
	k_{\parallel}$, $z\cosh\xi=\left(  \varrho c/\varepsilon\right)  \left\vert
	\mathbf{k}\right\vert $.} $z$, $\xi$, $\nu$,%
\begin{equation}
	z=\frac{c\varrho}{\varepsilon}\left\vert \mathbf{k}_{\perp}\right\vert
	,\ \ \xi=\frac{1}{2}\ln\left(  \frac{\left\vert \mathbf{k}\right\vert
		+k_{\parallel}}{\left\vert \mathbf{k}\right\vert -k_{\parallel}}\right)
	,\ \ \nu=\frac{\mathbf{k}_{\perp}\underline{\mathbf{u}}_{\perp}}{\varepsilon
	},\ \ \left\vert \mathbf{k}_{\perp}\right\vert \neq0\,,\label{s3.3}%
\end{equation}
to rewrite the phase $\Phi\left(  t^{\prime}\right)  $ as follows%
\begin{equation}
	\Phi\left(  \eta^{\prime}\right)  =z\sinh\left(  \eta^{\prime}-\xi\right)
	-\nu\eta^{\prime}+\mathcal{\tilde{C}}\ ,\label{s3.4}%
\end{equation}
where $\mathcal{\tilde{C}}=-\omega\left(  \underline{u}_{\Vert}/c\varepsilon
+t_{\mathrm{in}}\right)  -\mathbf{k}\underline{\mathbf{r}}+\nu
\ \mathrm{arc\sinh}\left(  \underline{u}_{\Vert}/c\varrho\right)
=\mathrm{const}$. Note that we used $\left\vert \mathbf{k}\right\vert $
instead of $k_{0}$ in Eq. (\ref{s3.3}) for convenience. As a result, the
complex function (\ref{s3.1}) admits the representation%
\begin{eqnarray}
	y_{\mathbf{k}\lambda}\left(  t,t_{\mathrm{in}}\right)  &=&-i\frac{qc}{2\pi}%
	\frac{e^{-ik_{0}ct_{\mathrm{in}}}e^{i\mathcal{\tilde{C}}}}{\varepsilon
		\sqrt{\hslash\omega}}\boldsymbol{\epsilon}_{\mathbf{k}\lambda}^{\ast
	}\mathbf{I}_{\nu}\left(  t,t_{\mathrm{in}}\right)\,,\nonumber\\
   \mathbf{I}_{\nu
	}\left(  t,t_{\mathrm{in}}\right)  &=&\left(  \frac{\underline{\mathbf{u}%
		}_{\perp}}{c}I_{\nu}^{\left(  1\right)  }\left(  t,t_{\mathrm{in}}\right)
	,\varrho I_{\nu}^{\left(  2\right)  }\left(  t,t_{\mathrm{in}}\right)
	\right)  \,,\label{s3.5}%
\end{eqnarray}
where%
\begin{align}
	&  I_{\nu}^{\left(  1\right)  }\left(  t,t_{\mathrm{in}}\right)  =\int%
	_{\eta_{\mathrm{in}}}^{\eta}e^{i\left[  z\sinh\left(  \eta^{\prime}%
		-\xi\right)  -\nu\eta^{\prime}\right]  }d\eta^{\prime},\ \ I_{\nu}^{\left(
		2\right)  }\left(  t,t_{\mathrm{in}}\right)  =\int_{\eta_{\mathrm{in}}}^{\eta
	}e^{i\left[  z\sinh\left(  \eta^{\prime}-\xi\right)  -\nu\eta^{\prime}\right]
	}\sinh\eta^{\prime}d\eta^{\prime}\,,\nonumber\\
	&  \eta=\eta\left(  t\right)  =\mathrm{arc}\sinh\left(  \varepsilon
	t/\varrho+\underline{u}_{\Vert}/\varrho c\right)  ,\ \ \eta_{\mathrm{in}%
	}\equiv\eta\left(  t_{\mathrm{in}}\right)  \,.\label{s3.6}%
\end{align}
By performing a supplementary change of variable $u^{\prime}=\eta^{\prime}%
-\xi$ allows us to express the above integrals in terms of an
\textquotedblleft incomplete\textquotedblright\ Macdonald function,%
\begin{equation}
	K_{i\nu}\left(  z;t,t_{\mathrm{in}}\right)  =\frac{e^{-\pi\nu/2}}{2}%
	\int_{u_{\mathrm{in}}}^{u}e^{i\phi\left(  u^{\prime}\right)  }du^{\prime
	},\ \ \phi\left(  u^{\prime}\right)  =z\sinh u^{\prime}-\nu u^{\prime
	}\,,\label{s3.7}%
\end{equation}
as follows%
\begin{eqnarray}
	  I_{\nu}^{\left(  1\right)  }\left(  t,t_{\mathrm{in}}\right)  &=&2e^{-i\nu
		\xi}e^{\pi\nu/2}K_{i\nu}\left(  z;t,t_{\mathrm{in}}\right)  \,,\nonumber\\
	  I_{\nu}^{\left(  2\right)  }\left(  t,t_{\mathrm{in}}\right)  &=&2e^{-i\nu
		\xi}e^{\pi\nu/2}\frac{\left\vert \mathbf{k}\right\vert }{\left\vert
		\mathbf{k}_{\perp}\right\vert }\left[  \frac{\nu}{z}\frac{k_{\parallel}%
	}{\left\vert \mathbf{k}\right\vert }K_{i\nu}\left(  z;t,t_{\mathrm{in}%
	}\right)  -iS_{i\nu}\left(  z;t,t_{\mathrm{in}}\right)  \right]
	\,,\nonumber\\
	  S_{i\nu}\left(  z;t,t_{\mathrm{in}}\right)  &=&K_{i\nu}^{\prime}\left(
	z;t,t_{\mathrm{in}}\right)  -\frac{1}{z}\frac{k_{\parallel}}{\left\vert
		\mathbf{k}\right\vert }\dot{K}_{i\nu}\left(  z;t,t_{\mathrm{in}}\right)
	\,,\label{s3.8}%
\end{eqnarray}
where $u=\eta-\xi$, $u_{\mathrm{in}}=\eta_{\mathrm{in}}-\xi$, and%
\begin{align}
	K_{i\nu}^{\prime}\left(  z;t,t_{\mathrm{in}}\right)   &  =\partial_{z}K_{i\nu
	}\left(  z;t,t_{\mathrm{in}}\right)  =\frac{i}{2}\left[  e^{-\pi}K_{i(\nu
		-1)}\left(  z;t,t_{\mathrm{in}}\right)  -e^{+\pi}K_{i(\nu+1)}\left(
	z;t,t_{\mathrm{in}}\right)  \right]  \ ,\nonumber\\
	\dot{K}_{i\nu}\left(  z;t,t_{\mathrm{in}}\right)   &  =\partial_{\xi}K_{i\nu
	}\left(  z;t,t_{\mathrm{in}}\right)  =\frac{e^{-\pi\nu/2}}{2}\left[
	e^{i\left(  z\sinh u_{\mathrm{in}}-\nu u_{\mathrm{in}}\right)  }-e^{i\left(
		z\sinh u-\nu u\right)  }\right]  \ .\label{s3.8b}%
\end{align}
It is important to mention that the integral (\ref{s3.7}) can be alternatively
expressed in terms of in complete cylindrical function
of the Bessel form $\epsilon_{\nu}(a,z)$ (see Ref. \cite{Agrest}),%
\begin{equation}
	K_{i\nu}\left(  z;t,t_{\mathrm{in}}\right)  =\frac{i\pi}{2}e^{-\pi\nu
		/2}\left[  \epsilon_{i\nu}(u,iz)-\epsilon_{i\nu}(u_{\mathrm{in}},iz)\right]
	,\ \ \epsilon_{\nu}(a,z)=\frac{1}{\pi i}\int_{0}^{a}e^{z\sinh t-\nu
		t}dt\ .\label{s3.8c}%
\end{equation}
For large values of the parameter $z$ and real values of $a$, the function
$\epsilon_{\nu}(a,z)$ behaves like%
\begin{align}
	\epsilon_{\nu}(a,z) &  =\frac{1}{\pi iz}\left(  \frac{e^{z\sinh a-\nu a}%
	}{\cosh a}-1\right)  +\frac{1}{\pi iz^{2}}\left[  \frac{e^{z\sinh a-\nu a}%
	}{\cosh^{2}a}\left(  \nu+\tanh a\right)  -\nu\right]  \nonumber\\
	&  +\frac{1}{\pi iz^{3}}\left[  \frac{e^{z\sinh a-\nu a}}{\cosh^{3}a}\left(
	\nu^{2}-1+3\nu\tanh a-3\tanh^{2}a\right)  -\left(  \nu^{2}-1\right)  \right]
	+O\left(  z^{-4}\right)  \ .\label{e1z}%
\end{align}

Finally, calculating the modulus square of Eq. (\ref{s3.5}) and summing the
result over the photon polarizations with the aid of the identities
(\ref{a.4b}), the total electromagnetic energy radiated by the particle
(\ref{f10b}) takes the form:%
\begin{equation}
	W\left(  t,t_{\mathrm{in}}\right)  =\left(  \frac{qc}{\varepsilon\pi}\right)
	^{2}\int e^{\pi\nu}\left\{  \left[  \left(  1-\frac{\nu^{2}}{z^{2}}\right)
	\varrho^{2}-1\right]  \left\vert K_{i\nu}\left(  z;t,t_{\mathrm{in}}\right)
	\right\vert ^{2}+\varrho^{2}\left\vert S_{i\nu}\left(  z;t,t_{\mathrm{in}%
	}\right)  \right\vert ^{2}\right\}  d\mathbf{k}\,.\label{s3.9}%
\end{equation}
This compact expression corresponds to a generalization of the classical
differential energy due to its dependence on time; cf. Eq. (16) in Ref.
\cite{NikRit69}. As it can be seen, the difference between Eqs. (\ref{s3.9})
and the classical result lies in the definition of the integral (\ref{s3.7}),
which, contrary to the Macdonald function,%
\begin{equation}
	K_{i\nu}\left(  z\right)  =\frac{e^{-\pi\nu/2}}{2}\int_{-\infty}^{\infty} e^{i\left(  z\sinh
		s-\nu s\right)  }ds,\ \ K_{iv}^{\prime}\left(  z\right)  =\frac{d}{dz}K_{i\nu
	}\left(  z\right)  \,,\label{s3.9b}%
\end{equation}
has a finite integration range.

As discussed before, the time dependence is a consequence of the fact that the
electromagnetic radiation in the semiclassical formulation forms within the
quantum transition interval $\Delta t=t-t_{\mathrm{in}}$. That is why the
electromagnetic energy (\ref{s3.9}) (and the energy rate, see Eq.
(\ref{s3.10})\textrm{ }below) are time-dependent. In classical
electrodynamics, on the other hand, the Macdonald function (\ref{s3.9b})
results from the application of Poynting's theorem to the computation of the
energy \cite{NikRit69}\footnote{In classical theory, there is the possibility
	of defining the time-dependent electromagnetic energy by assuming that the
	current is exposed to the external field over a finite time interval
	\cite{Jacks99}. However, such time-dependent energy differs from the one
	obtained within the semiclassical formulation discussed here.}. As a result,
the energy is time-independent. As discussed in Sec. \ref{S2}, we may recover
the classical result setting $t\rightarrow+\infty$, $t_{\mathrm{in}%
}\rightarrow-\infty$. However, implementing this limit directly in
(\ref{s3.7}) requires adding a converging factor to the integrand in order to
eliminate oscillatory contributions from $\dot{K}_{i\nu}\left(
z;t,t_{\mathrm{in}}\right)  $ at remote time limits. For example, a
possibility could be adding the convergence factor\textrm{ }$\exp\left(
-\epsilon\left\vert u^{\prime}\right\vert \right)  $ to the integral
(\ref{s3.7}),%
\[
\lim_{\epsilon\rightarrow0}\int_{u_{\mathrm{in}}}^{u}e^{-\epsilon\left\vert
	u^{\prime}\right\vert }e^{i\phi\left(  u^{\prime}\right)  }du^{\prime
}\,,\ \ \epsilon>0\,,
\]
and performing the limit at the end of calculations. In this sense, $\left.
\dot{K}_{i\nu}\left(  z;t,t_{\mathrm{in}}\right)  \right\vert _{t_{\mathrm{in}%
	}\rightarrow-\infty,t\rightarrow+\infty}$ vanishes, the integral (\ref{s3.7})
coincides with the Macdonald function (\ref{s3.9b}) and the energy (\ref{s3.9})
agrees with the classical result, $W\left(  +\infty,-\infty\right)  \equiv
W_{\mathrm{cl}}$ (cf. Eq. (16) in Ref. \cite{NikRit69}). Another possibility
could be expressing the differential energy (\ref{s3.9}) in terms of the
$I$-integrals (\ref{s3.6}) and implementing the limits $t_{\mathrm{in}%
}\rightarrow-\infty$, $t\rightarrow+\infty$ before the change of variables
$u^{\prime}=\eta^{\prime}-\xi$. This alternative leads directly to the
classical result found in Ref. \cite{NikRit69}.

\begin{figure}[h]
	\centering
	\includegraphics[width=0.6\textwidth]{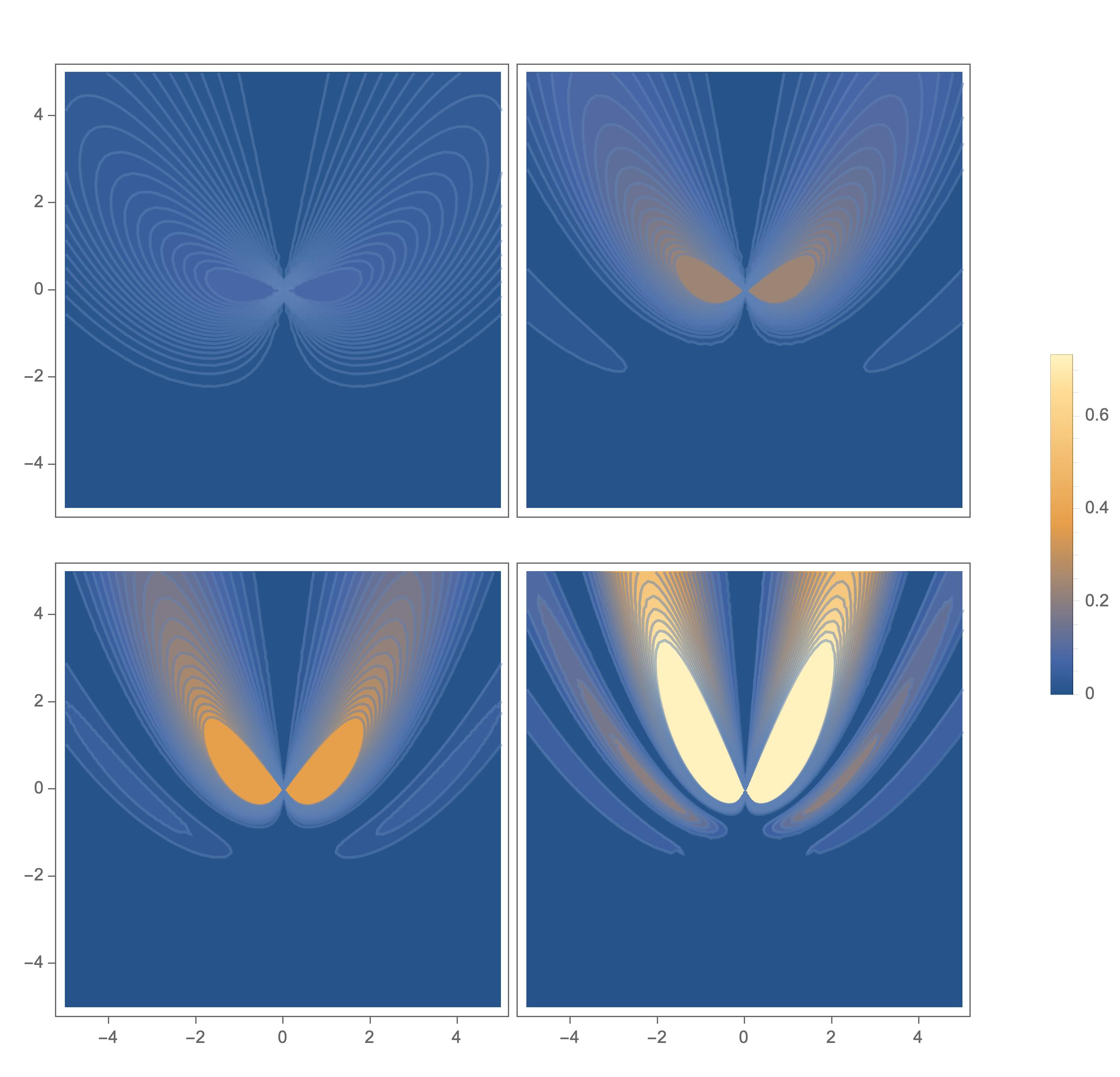}
	\caption{Spectral-angular distribution of the total radiation energy
		(\ref{nn01}) as a function of $t$, $k_{x}$ (horizontal axes) and
		$k_{\parallel}$ (vertical axes) for $c/\varepsilon=0.1$. In a) $\eta=3$, b)
		$\eta=3.6$, c) $\eta=4$, and d) $\eta=4.5$. Moreover, in all plots we chose
		$t_{\mathrm{in}}$ such that $\eta_{\mathrm{in}}=0$ for convenience.}
	\label{Fig1}
\end{figure}

Next, we examine specific representations for the electromagnetic energy
radiated by the particle (\ref{s3.9}). If the particle moves parallel to the
external field--that is, its trajectory is subjected to the initial condition
$\underline{\mathbf{v}}_{\perp}=\mathbf{0}$--then $\varrho=1$, $\nu=0$, and
the energy (\ref{s3.9}) admits the form%
\begin{equation}
	\left.  W\left(  t,t_{\mathrm{in}}\right)  \right\vert _{\underline{\mathbf{v}%
		}_{\perp}=\mathbf{0}}=\left(  \frac{qc}{\pi\varepsilon}\right)  ^{2}%
	\int\left\vert S_{0}\left(  z;t,t_{\mathrm{in}}\right)  \right\vert
	^{2}d\mathbf{k}=\left(  \frac{qc}{2\pi\varepsilon}\right)  ^{2}\int%
	\frac{\mathbf{k}_{\perp}^{2}}{\mathbf{k}^{2}}\left\vert I_{0}^{\left(
		2\right)  }\left(  \mathbf{k};t,t_{\mathrm{in}}\right)  \right\vert
	^{2}d\mathbf{k}\,, \label{n01}%
\end{equation}
where $S_{0}\left(  z;t,t_{\mathrm{in}}\right)  $ is given by Eq. (\ref{s3.8})
and%
\begin{eqnarray}
	  \dot{K}_{0}\left(  z;t,t_{\mathrm{in}}\right)  &=&\left\{
	\begin{array}
		[c]{cl}%
		\left(  e^{iz\sinh u_{\mathrm{in}}}-e^{iz\sinh u}\right)  /2 & \mathrm{if}%
		\ -\infty<t_{\mathrm{in}}<t<+\infty\ ,\\
		0 & \mathrm{if\ }t=-t_{\mathrm{in}}=+\infty\ .
	\end{array}
	\right. \label{nn03}\\
	  I_{0}^{\left(  2\right)  }\left(  \mathbf{k};t,t_{\mathrm{in}}\right)
	&=&\int_{\eta_{\mathrm{in}}}^{\eta}\exp\left[  i\frac{c}{\varepsilon}\left(
	\left\vert \mathbf{k}\right\vert \sinh\eta^{\prime}-k_{\parallel}\cosh
	\eta^{\prime}\right)  \right]  \sinh\eta^{\prime}d\eta^{\prime}\,.
	\label{nn04}%
\end{eqnarray}
The corresponding spectral-angular distribution of the total electromagnetic
energy radiated by the particle reads:%
\begin{equation}
	\left.  \frac{d^{3}W\left(  t,t_{\mathrm{in}}\right)  }{dk_{x}dk_{y}dk_{z}%
	}\right\vert _{\underline{\mathbf{v}}_{\perp}=\mathbf{0}}=\left(  \frac
	{qc}{\varepsilon\pi}\right)  ^{2}\left\vert S_{0}\left(  z;t,t_{\mathrm{in}%
	}\right)  \right\vert ^{2}=\left(  \frac{qc}{2\pi\varepsilon}\right)
	^{2}\frac{\mathbf{k}_{\perp}^{2}}{\mathbf{k}^{2}}\left\vert I_{0}^{\left(
		2\right)  }\left(  \mathbf{k};t,t_{\mathrm{in}}\right)  \right\vert ^{2}\,.
	\label{nn01}%
\end{equation}
To further explore the angular dependence, we represent the spectral-angular
total energy distribution (\ref{nn01}) in spherical coordinates%
\begin{eqnarray}
	  \frac{d^{2}W\left(  t,t_{\mathrm{in}}\right)  }{k_{0}^{2}dk_{0}d\Omega
	}&=&\left(  \frac{qc}{2\pi\varepsilon}\right)  ^{2}\sin^{2}\theta\ \left\vert
	I_{0}^{\left(  2\right)  }\left(  \theta;t,t_{\mathrm{in}}\right)
	)\right\vert ^{2},\ \ d\Omega=\sin\theta d\theta d\varphi\ ,\nonumber\\
	I_{0}^{\left(  2\right)  }\left(  \theta;t,t_{\mathrm{in}}\right)
	&=&\int_{\eta_{\mathrm{in}}}^{\eta}\sinh\eta^{\prime}\exp\left[  i\Lambda\left(
	\sinh\eta^{\prime}-\cos\theta\cosh\eta^{\prime}\right)  \right]  d\eta
	^{\prime}\,,\ \ \Lambda=\frac{c^{2}k_{0}}{a}\ . \label{n02}%
\end{eqnarray}
It must be noted that the expression (\ref{n02}) depends exclusively on
$k_{0}=\sqrt{k_{x}^{2}+k_{y}^{2}+k_{z}^{2}}$, the angle $\theta$, and not on
the polar angle $\varphi=\arctan k_{y}/k_{x}$, for instance. Therefore, for
illustrative purposes we present in Fig. \ref{Fig1} the spectral-angular
distribution (\ref{nn01}) as a function of $k_{x}$ (horizontal axes),
$k_{\parallel}=k_{z}$ (vertical axes) and for specific times, assuming
$c/\varepsilon=0.1$, $q=2$, and $k_{y}=0$.%

\begin{figure}[h]
	\centering
	\includegraphics[width=0.6\textwidth]{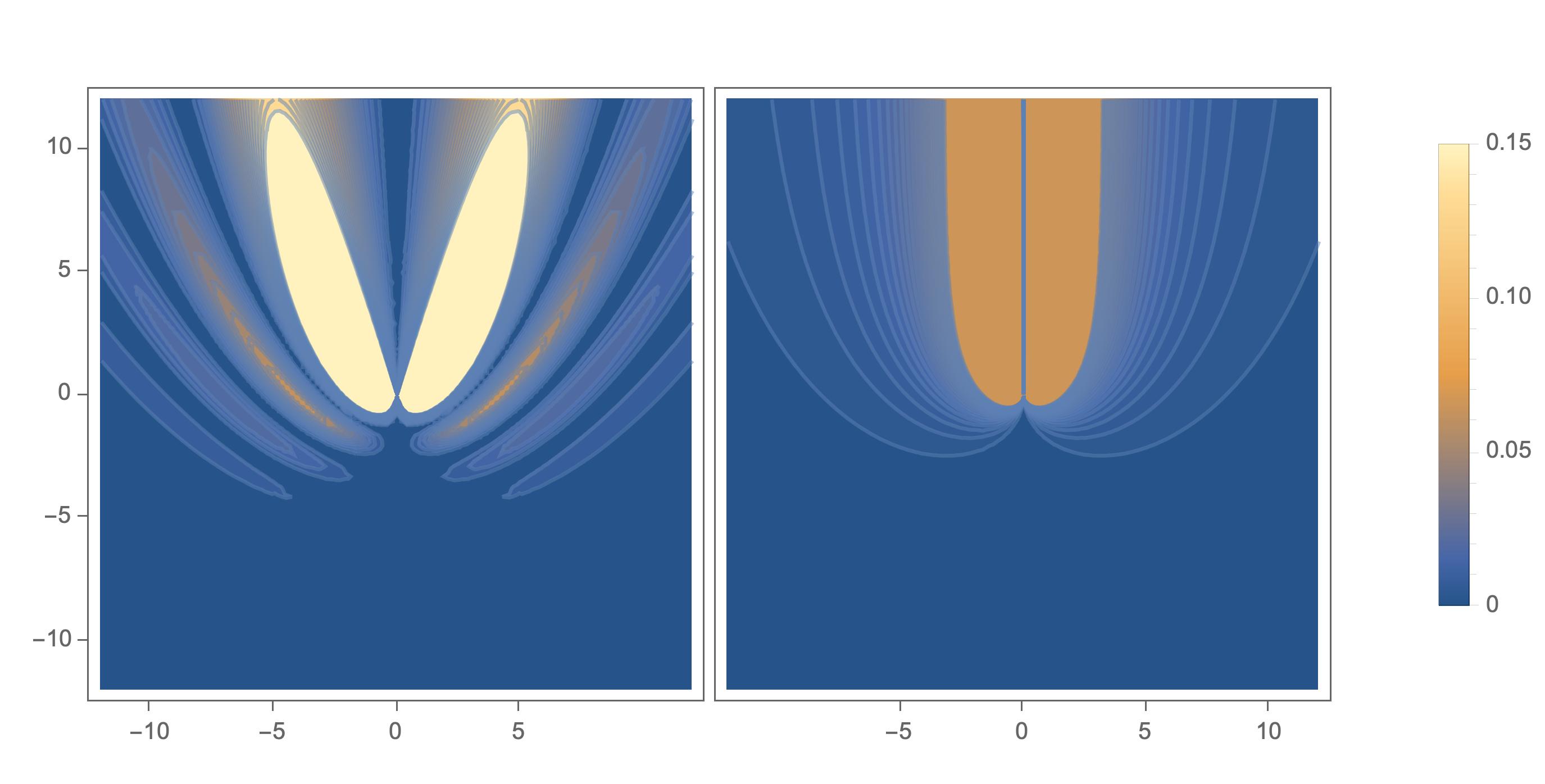}
	\caption{Spectral-angular distribution of the total radiation energy as a
		function of $t$, $k_{x}$ (horizontal axes), and $k_{\parallel}$ (vertical
		axes) for $\eta_{\mathrm{in}}=0$, $\eta=3.8$. The left panel (a) illustrates
		the exact expression (\ref{n02}) while the right panel (b) illustrates its
		asymptotic form, given by Eq. (\ref{n04}).}
	\label{Fig2}
\end{figure}

Consider the case when the parameter $\Lambda$ is large. In this case, the
integral $I_{0}^{\left(  2\right)  }\left(  \theta;t,t_{\mathrm{in}}\right)  $
can be estimated by the stationary phase method:%
\begin{eqnarray}
	I_{0}^{\left(  2\right)  }\left(  \theta;t,t_{\mathrm{in}}\right)
	&=&I_{\mathrm{as}}(\theta;t,t_{\mathrm{in}})+O\left(  \frac{1}{\Lambda^{2}%
	}\right)  ,\ \nonumber\\
 I_{\mathrm{as}}(\theta;t,t_{\mathrm{in}})&=&\frac{1}{i\Lambda
	}\left.  \frac{\sinh\eta^{\prime}}{\cosh\eta^{\prime}-\cos\theta\sinh
		\eta^{\prime}}e^{i\Lambda\left(  \sinh\eta^{\prime}-\cos\theta\cosh
		\eta^{\prime}\right)  }\right\vert _{\eta_{\mathrm{in}}}^{\eta}\,. \label{n03}%
\end{eqnarray}
It should be noted that the leading-order approximation obtained via the
stationary-phase method (\ref{n03}) could be alternatively obtained from the
exact expression (\ref{s3.9}) by taking into account the first term in
(\ref{e1z}) and using the Eqs. (\ref{s3.6})--(\ref{s3.8b}). Setting
$\eta_{\mathrm{in}}=0$, we obtain%
\begin{equation}
	\left.  \frac{d^{2}W\left(  t,t_{\mathrm{in}}\right)  }{k_{0}^{2}dk_{0}%
		d\Omega}\right\vert _{\eta_{\mathrm{in}}=0}=\left(  \frac{q}{2\pi k_{0}}%
	\frac{\sin\theta\sinh\eta}{\cosh\eta-\cos\theta\sinh\eta}\right)
	^{2}+O\left(  \frac{1}{\Lambda^{4}}\right)  \ . \label{n04}%
\end{equation}
For large $\Lambda$, (\ref{n02}) is a good approximation for the exact
expression (\ref{n02}). The angle for which the spectral-angular distribution
is maximum at large $\Lambda$ can be estimated from (\ref{n04}):%
\[
\theta_{\max}\simeq\pm\arccos\tanh\eta\ .
\]
For small values of $\Lambda$, the integral $I_{0}^{\left(  2\right)  }\left(
\theta;t,t_{\mathrm{in}}\right)  $ oscillates around $I_{\mathrm{as}}%
(\theta;t,t_{\mathrm{in}})$ and ensures the presence of bands in the
spectral-angular distribution, as illustrated in Fig. \ref{Fig2}. In this
figure, we present the spectral-angular distribution (\ref{n02}) as a function
of $k_{x}$ (horizontal axes), $k_{\parallel}$ (vertical axes) for $k_{y}=0$,
$\eta_{\mathrm{in}}=0$, $\eta=3.8$, and $c/\varepsilon=0.1$, $q=2$. The left
panel corresponds to the exact expression (\ref{n02}) while the right panel
corresponds to the asymptotic approximation truncated to the leading order,
given by Eq. (\ref{n04}).%

To conclude this section, we comment on the absence of divergences associated
with the computation of the total energy radiated by the particle
(\ref{s3.9}). In work \cite{NikRit69}, Nikishov and Ritus pointed out that the
total energy is divergent because the particle radiates at a constant rate
during its motion. Thus, if the electric field acts on the particle for an
infinite time interval, the energy emitted by the particle diverges.
Technically, the divergence stems from the classical spectrum being independent of
$k_{\parallel}$. Therefore, the integral of the differential energy diverges
linearly with $k_{\parallel}$. Moreover, the spectrum features
logarithmically-devergent contributions at $\left\vert \mathbf{k}_{\perp
}\right\vert =0$ due to the behavior of the Macdonald functions (\ref{s3.9b})
with small argument: for the linear motion, for instance, $K_{0}\left(
z\right)  =-\gamma-\ln\left(  z/2\right)  +O\left(  z\log\left(  z\right)
\right)  $, where $\gamma\approx0.577$ is Euler's constant. Since the
differential energy is proportional to $K_{0}^{\prime}\left(  z\right)  $, the
integrand is singular at $\left\vert \mathbf{k}_{\perp}\right\vert =0$. As
noticed by Nikishov and Ritus, the above mentioned problems are linked to the
perpetual duration of the electric field and it is clear that they can be
avoided introducing a regularization in time; e.g., assuming that the electric
field switches-on and -off at specific times. In the semiclassical approach,
we observe that the energy is finite because the radiation is formed within
the quantum transition interval $t-t_{\mathrm{in}}$. This \textquotedblleft
radiation formation interval\textquotedblright\ naturally introduces a
regularization to the problem and eliminates divergences mentioned above. To
see it more clearly, it is enough inspecting Eq. (\ref{n01}). Contrary to the
classical energy, the semiclassical expressions (\ref{s3.9}), (\ref{n01})
depends explicitly on $k_{\parallel}$; thus, linear divergences in
$k_{\parallel}$ do not arise in the computation of (\ref{n01}). Secondly, from
Eq. (\ref{s3.8c}) and the power series representation of in complete
cylindrical function of the Bessel form \cite{Agrest},%
\begin{equation}
	\epsilon_{i\nu}\left(  u,iz\right)  =\sum_{n=0}^{\infty}\frac{R_{n}}%
	{n!}\left(  iz\right)  ^{n}\,,\ \ R_{n}=\frac{2^{-n}}{i\pi}\sum_{l=0}^{n}%
	\frac{\left(  -1\right)  ^{l}\Gamma\left(  n+1\right)  }{\Gamma\left(
		l+1\right)  \Gamma\left(  n-l+1\right)  }\frac{e^{\left(  n-2l-i\nu\right)
			u}-1}{n-2l-i\nu}\,, \label{n04b}%
\end{equation}
we see that the incomplete Macdonald function (\ref{s3.7}) does not feature
logarithmic terms. In particular, the integral $K_{0}\left(
z;t,t_{\mathrm{in}}\right)  $ admits the following expansion:%
\begin{equation}
	K_{0}\left(  z;t,t_{\mathrm{in}}\right)  =\frac{u-u_{\mathrm{in}}}{2}+\frac
	{1}{2}\sum_{n=1}^{\infty}\frac{\left(  iz\right)  ^{n}}{2^{n}}\sum_{l=1}%
	^{n}\frac{\left(  -1\right)  ^{l}}{\Gamma\left(  l+1\right)  \Gamma\left(
		n-l+1\right)  }\frac{e^{\left(  n-2l\right)  u}-e^{\left(  n-2l\right)
			u_{\mathrm{in}}}}{n-2l}\,. \label{n04c}%
\end{equation}
Therefore, the integrand in (\ref{s3.9}) and in (\ref{n01}) are regular at
$\left\vert \mathbf{k}_{\perp}\right\vert =0$.

\subsection{Total energy rate\label{Sec3.2}}

In this subsection we focus the attention to the total energy rate, discussed
in subsection \ref{S2.2}. In terms of the integrals (\ref{s3.6}), the total
energy rate (\ref{f14}) can be preliminarily presented in the form%
\begin{align*}
	  \ w\left(  t,t_{\mathrm{in}}\right)  &=\left(  \frac{qc}{\pi}\right)
	^{2}\int\frac{1}{2}\operatorname{Re}\left\{  \frac{e^{-i\mathcal{\tilde{C}}%
		}e^{i\Phi\left(  t\right)  }}{\varepsilon}\boldsymbol{\beta}\left(  t\right)
	\mathbf{I}_{\nu}^{\ast}\left(  z;t,t_{\mathrm{in}}\right)  \right.  \\
	&  -\left.  e^{-i\mathcal{\tilde{C}}}e^{i\Phi\left(  t\right)  }%
	\frac{\boldsymbol{\beta}\left(  t\right)  \mathbf{k}}{\mathbf{k}^{2}}\left[
	\frac{\nu}{c}I_{\nu}^{\left(  1\right)  \ast}\left(  z;t,t_{\mathrm{in}%
	}\right)  +\frac{\varrho}{\varepsilon}k_{\parallel}I_{\nu}^{\left(  2\right)
		\ast}\left(  z;t,t_{\mathrm{in}}\right)  \right]  \right\}  d\mathbf{k}\,.
\end{align*}
Performing the change of variables $t=-\underline{u}_{\Vert}/\varepsilon
c+\left(  \varrho/\varepsilon\right)  \sinh\left(  u+\xi\right)  $ and
expressing the above integrals in terms of\ (\ref{s3.7}) through the
identities (\ref{s3.8}) the energy rate takes the form%
\begin{eqnarray}
	  w\left(  t,t_{\mathrm{in}}\right)  &=&\left(  \frac{qc}{\pi}\right)
	^{2}\frac{1}{\varepsilon\varrho}\int\frac{e^{\pi\nu/2}}{\cosh\eta}\left\{
	\left[  \left(  1-\frac{\nu^{2}}{z^{2}}\right)  \varrho^{2}-1\right]
	\operatorname{Re}\left[  e^{i\phi\left(  u\right)  }K_{i\nu}^{\ast}\left(
	z;t,t_{\mathrm{in}}\right)  \right]  \right.  \nonumber\\
	&+&\left.  \frac{\left\vert \mathbf{k}_{\perp}\right\vert }{\left\vert
		\mathbf{k}\right\vert }\left(  \sinh\eta-\frac{\nu}{z}\frac{k_{\parallel}%
	}{\left\vert \mathbf{k}_{\perp}\right\vert }\right)  \varrho^{2}%
	\operatorname{Re}\left[  ie^{i\phi\left(  u\right)  }S_{i\nu}^{\ast}\left(
	z;t,t_{\mathrm{in}}\right)  \right]  \right\}  d\mathbf{k}\,.\label{s3.10}%
\end{eqnarray}
This equation is our main result. It expresses the electromagnetic energy
radiated by the particle within the quantum transition interval $\Delta
t=t-t_{\mathrm{in}}$. Similarly to the total energy (\ref{s3.9}), Eq.
(\ref{s3.10}) corresponds to a generalization of the classical energy rate
radiated by the particle accelerated by the electric field. To simplify
calculations and the comparison with the classical result, we perform the
change of variables $\left(  k_{x},k_{y},k_{z}\right)  \rightarrow\left(
k_{\parallel},\left\vert \mathbf{k}_{\perp}\right\vert ,\vartheta\right)  $,%
\begin{equation}
	k_{\parallel}=k_{z},\ \ \mathbf{k}_{\perp}^{2}=k_{x}^{2}+k_{y}^{2}%
	,\ \ \vartheta=\arccos\left(  \frac{\underline{\mathbf{P}}_{\perp}%
		\mathbf{k}_{\perp}}{\left\vert \underline{\mathbf{P}}_{\perp}\right\vert
		\left\vert \mathbf{k}_{\perp}\right\vert }\right)  \,,\ \ d\mathbf{k}%
	=\left\vert \mathbf{k}_{\perp}\right\vert d\left\vert \mathbf{k}_{\perp
	}\right\vert dk_{\parallel}d\vartheta\,,\label{s3.12}%
\end{equation}
and choose $t_{\mathrm{in}}=-\infty$ to find that the energy rate admits form%
\begin{eqnarray}
	 w\left(  t\right)  &=&\left(  \frac{qc}{\pi}\right)  ^{2}\frac{\varrho
	}{\varepsilon}\int\frac{e^{\pi\nu/2}}{\cosh\eta}\left\{  \frac{1}{\varrho^{2}%
	}\left[  \left(  1-\frac{\nu^{2}}{z^{2}}\right)  \varrho^{2}-1\right]
	\operatorname{Re}\left[  e^{i\phi\left(  u\right)  }K_{i\nu}^{\ast}\left(
	z;t\right)  \right]  \right.  \nonumber\\
	&+&\left.  \frac{\left\vert \mathbf{k}_{\perp}\right\vert }{\left\vert
		\mathbf{k}\right\vert }\left(  \sinh\eta-\frac{\nu}{z}\frac{k_{\parallel}%
	}{\left\vert \mathbf{k}_{\perp}\right\vert }\right)  \operatorname{Im}\left[
	e^{-i\phi\left(  u\right)  }K_{i\nu}^{\prime}\left(  z;t\right)  \right]
	\right\}  d\mathbf{k}\,.\label{s3.11}%
\end{eqnarray}
Here, $w\left(  t\right)  \equiv w\left(  t,-\infty\right)  $ and%
\begin{equation}
	K_{i\nu}\left(  z;t\right)  \equiv K_{i\nu}\left(  z;t,-\infty\right)
	=\frac{e^{-\pi\nu/2}}{2}\int_{-\infty}^{u}e^{i\phi\left(  u^{\prime}\right)
	}du^{\prime}\,.\label{s3.11b}%
\end{equation}

In the large time limit $t\rightarrow+\infty$, we notice that only one term in
(\ref{s3.11}) contributes to the total energy rate, namely%
\begin{eqnarray}
	w\equiv\lim_{t\rightarrow\infty}w\left(  t\right) &=& 
	-\left(  \frac{qc}{\pi
	}\right)  ^{2}\frac{\varrho}{\varepsilon}\int_{0}^{2\pi}d\vartheta\int%
	_{0}^{\infty}d\left\vert \mathbf{k}_{\perp}\right\vert e^{\pi\nu/2}K_{i\nu
	}^{\prime}\left(  z\right)  \nonumber\\
    &\times&\int_{-\infty}^{+\infty}\frac{\mathbf{k}_{\perp
		}^{2}}{\sqrt{\mathbf{k}_{\perp}^{2}+k_{\parallel}^{2}}}\lim_{t\rightarrow
		\infty}\left[  \sin\phi\left(  u\right)  \right]  dk_{\parallel}\,.
	\label{s3.13}%
\end{eqnarray}

The computation of the rate simplifies considerably if we restrict ourselves
to the case where the particle is subjected to the initial condition
$\underline{\mathbf{v}}_{\perp}=\mathbf{0}$. In this case, $\varrho=1$,
$\nu=0$, and the energy rate (\ref{s3.11}) assumes the form%
\begin{align}
	&\left.  w\left(  t\right)  \right\vert _{\underline{\mathbf{v}}_{\perp
		}=\mathbf{0}}=\frac{2\left(  qc\right)  ^{2}}{\pi\varepsilon}\tanh\eta\nonumber\\
	&\times\int%
	_{0}^{\infty}d\left\vert \mathbf{k}_{\perp}\right\vert \mathbf{k}_{\perp}%
	^{2}\int\frac{\cos\left(  z\sinh u\right)  \operatorname{Im}K_{0}^{\prime
		}\left(  z;t\right)  -\sin\left(  z\sinh u\right)  \operatorname{Re}%
		K_{0}^{\prime}\left(  z;t\right)  }{\sqrt{\mathbf{k}_{\perp}^{2}+k_{\parallel
			}^{2}}}dk_{\parallel}\,. \label{s3.13b}%
\end{align}
In the limit $t\rightarrow+\infty$, the integral (\ref{s3.11b}) becomes the
Macdonald function (\ref{s3.9b}) and
\begin{equation*}
	\operatorname{Im}K_{0}^{\prime}\left(z;+\infty\right)=0,\quad \operatorname{Re}K_{0}^{\prime}\left(  z;+\infty\right)=
	K_{0}^{\prime}\left(z\right)\,.
\end{equation*}
As a result,
\begin{eqnarray}
	\left.  w\right\vert _{\underline{\mathbf{v}}_{\perp}=\mathbf{0}}%
	&=&\lim_{t\rightarrow\infty}\left.  w\left(  t\right)  \right\vert
	_{\underline{\mathbf{v}}_{\perp}
		=\mathbf{0}}\nonumber\\
	&=&-\frac{2\left(  qc\right)  ^{2}%
	}{\pi\varepsilon}\int_{0}^{\infty}d\left\vert \mathbf{k}_{\perp}\right\vert
	K_{0}^{\prime}\left(  z\right)  \int_{-\infty}^{+\infty}\frac{\mathbf{k}%
		_{\perp}^{2}}{\sqrt{\mathbf{k}_{\perp}^{2}+k_{\parallel}^{2}}}\lim
	_{t\rightarrow\infty}\left[  \sin\left(  z\sinh u\right)  \right]
	dk_{\parallel}\,. \label{s3.14}%
\end{eqnarray}
For large times, the above limit has the form%
\begin{eqnarray*}
	  \lim_{t\rightarrow\infty}\sin\left(  z\sinh u\right)  &=&\lim_{t\rightarrow
		\infty}\sin\left[  \frac{c }{\varepsilon}\left(  \left\vert
	\mathbf{k}\right\vert ^{2}\sinh\eta-k_{\parallel}^{2}\cosh\eta\right)  \right]
	\\
	&=&\lim_{t\rightarrow\infty}\sin\left[  \frac{c }{\varepsilon}\left(
	\left\vert \mathbf{k}\right\vert ^{2}-k_{\parallel}^{2}\right)  \sinh
	\eta\right]\\
	&=&\lim_{t\rightarrow\infty}\sin\left[  \left(  \sqrt{\left(
		\varepsilon z\right)  ^{2}+\left(  ck_{\parallel}\right)  ^{2}}-ck_{\parallel
	}\right)  t\right]  \ .
\end{eqnarray*}
Thus, performing a supplementary change of variables $s=\sqrt{\left(
	\varepsilon z\right)  ^{2}+\left(  ck_{\parallel}\right)  ^{2}}-ck_{\parallel
}$ and using the identity $K_{0}^{\prime}\left(  z\right)  =-K_{1}\left(
z\right)  $ we finally obtain%
\begin{align}
	\left.  w\right\vert _{\underline{\mathbf{v}}_{\perp}=\mathbf{0}}  &
	=-\frac{2q^{2}\varepsilon^{2}}{\pi}\int_{0}^{\infty}K_{1}\left(  z\right)
	z^{2}dz\,\lim_{t\rightarrow\infty}\int_{\infty}^{0}\frac{\sin\left(
		st\right)  ds}{cs}\nonumber\\
	&  =\frac{q^{2}\varepsilon^{2}}{c }\int_{0}^{\infty}K_{1}\left(
	z\right)  z^{2}dz=2\frac{q^{2}}{c^{3}}a^{2}\,,\ \ a=\frac{1}{m}\frac
	{dP_{\parallel}\left(  t\right)  }{dt}=\frac{qE}{m}\,. \label{s3.15}%
\end{align}

Except by a factor of $1/3$, this result coincides with Larmor's formula for
the total energy rate radiated by an uniformily accelerated charged particle
\cite{Larmor1987,Jacks99}. The absence of this factor was also pointed out in
the framework of the classical theory by Nikishov and Ritus in Ref.
\cite{NikRit69}. However, in their work, this coefficient was somewhat hidden
by the presence of an extra factor of $3\pi^{2}/32$ resulting from the
integration of the Macdonald function $K_{1}\left(  z\right)  $ squared,
$\int_{0}^{\infty}K_{1}^{2}\left(  z\right)  z^{2}dz=3\pi^{2}/32$. The latter
integral arose in the computation of the total (classical) energy radiated by
the particle after relating the radiation formation interval $dt_{\mathrm{rf}%
}$ with the increment of the radiation's longitudinal wave number
$dk_{\parallel}$ produced within this interval, namely $dt_{\mathrm{rf}%
}=\left(  \varrho/\varepsilon\left\vert \mathbf{k}_{\perp}\right\vert \right)
dk_{\parallel}$. This identification enabled them to derive an effective
expression for the total energy rate directly from the total energy, which is
replicated here as follows: performing the change of variables (\ref{s3.12})
in the classical differential energy given by Eq. (16) in \cite{NikRit69} and
replacing $dk_{\parallel}$ by $dt_{\mathrm{rf}}$ according to relation above,
Nikishov and Ritus obtained the differential energy rate:%
\begin{equation}
	dw_{\mathrm{cl}}=\frac{dW_{\mathrm{cl}}}{dt_{\mathrm{rf}}}=\left(
	\frac{q\varepsilon}{\pi\varrho^{2}}\right)  ^{2}\frac{e^{\pi v}}{c}\left\{
	\left[  \left(  1-\frac{\nu^{2}}{z^{2}}\right)  \varrho^{2}-1\right]
	K_{iv}^{2}\left(  z\right)  +\varrho^{2}K_{iv}^{\prime2}\left(  z\right)
	\right\}  z^{2}dzd\vartheta\,. \label{s3.16}%
\end{equation}
Then, restricting to the linear motion ($\underline{\mathbf{v}}_{\perp
}=\mathbf{0}$), integrating over $\vartheta$ and $z$ they calculated the total
energy rate%
\begin{equation}
	\left.  w_{\mathrm{cl}}\right\vert _{\underline{\mathbf{v}}_{\perp}%
		=\mathbf{0}}=\frac{1}{c}\left(  \frac{q\varepsilon}{\pi}\right)  ^{2}\int%
	_{0}^{2\pi}d\vartheta\int_{0}^{\infty}K_{0}^{\prime2}\left(  z\right)
	z^{2}dz=\frac{3\pi}{32}\times\left.  w\right\vert _{\underline{\mathbf{v}%
		}_{\perp}=\mathbf{0}}\,, \label{s3.17}%
\end{equation}
and concluded that the result differs from Larmor's by the factor $9\pi/32$.

\begin{figure}[h]
	\centering
	\includegraphics[width=0.6\textwidth]{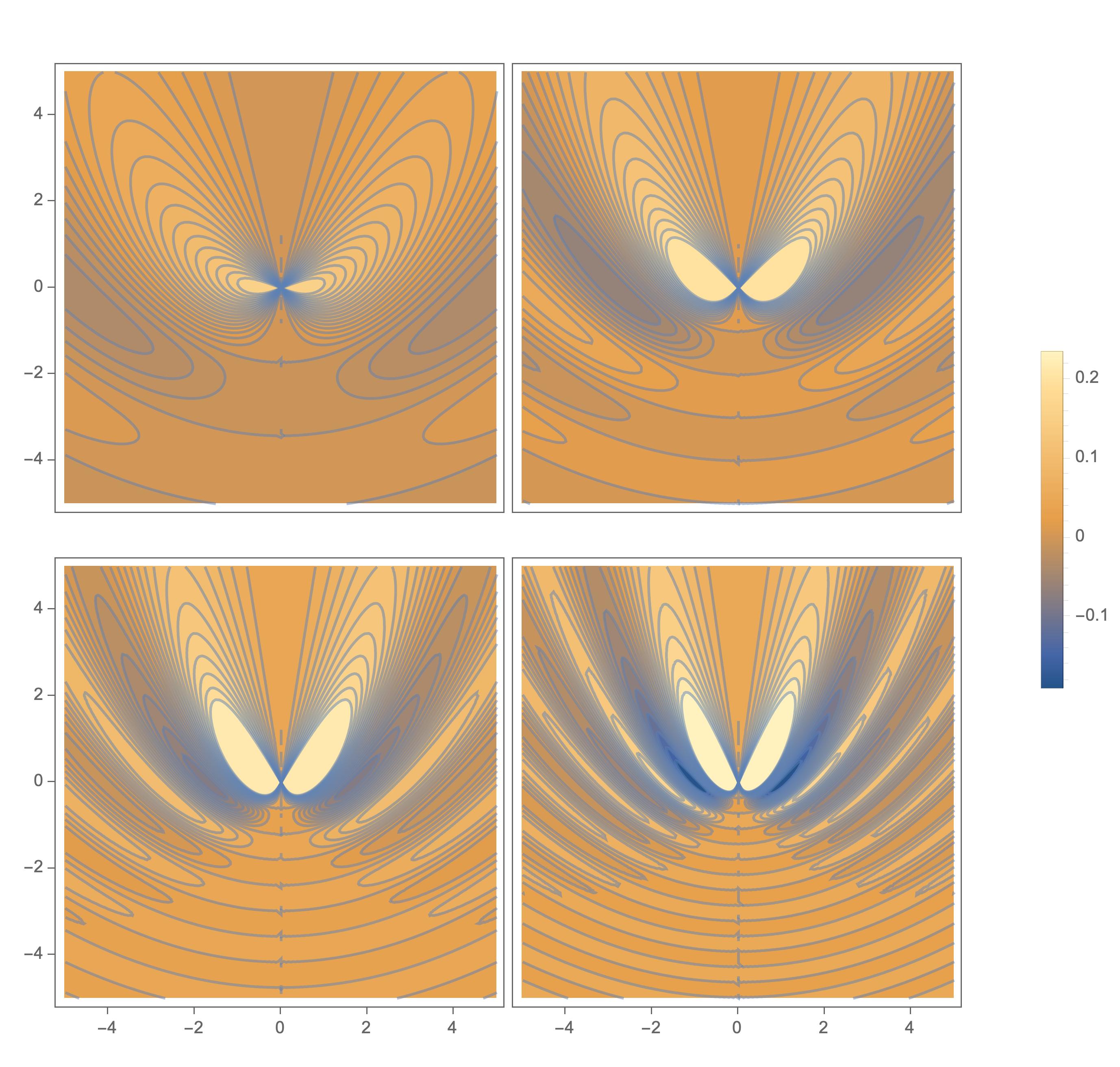}
	\caption{Spectral-angular distribution (\ref{n05}) as a function of $k_{x}$
		(horizontal axes), $k_{\parallel}$ (vertical axes) for $\varepsilon/c=0.1$. In
		a) $\eta=3$, b) $\eta=3.5$, c) $\eta=4$, and d) $\eta=4.5$.}
	\label{Fig3}
\end{figure}

Finally, it is worth inspecting the spectral distribution for the energy rate
in the case where the particle moves parallel to the field, i.e. when its
inicial velocity perpendicular to the field is zero $\underline{\mathbf{v}%
}_{\perp}=\mathbf{0}$. The corresponding expression can be either obtained
from Eq. (\ref{s3.10}) or by differentiating Eq. (\ref{n01}) with respect to
time:%
\begin{eqnarray}
	\left.  w\left(  t,t_{\mathrm{in}}\right)  \right\vert _{\underline{\mathbf{v}%
		}_{\perp}=\mathbf{0}}&=&\frac{1}{2\varepsilon}\left(  \frac{qc}{\pi}\right)
	^{2}\tanh\eta\nonumber\\
	&\times&\operatorname{Re}\int\frac{\mathbf{k}_{\perp}^{2}}%
	{\mathbf{k}^{2}}I_{0}^{\left(  2\right)  }\left(  \mathbf{k};t,t_{\mathrm{in}%
	}\right)  \exp\left[  -\frac{ic}{\varepsilon}\left(  \left\vert \mathbf{k}%
	\right\vert \sinh\eta-k_{\parallel}\cosh\eta\right)  \right]  d\mathbf{k}\,.
	\label{s3.20}%
\end{eqnarray}
Therefore, the corresponding spectral-angular distribution of the total energy
rate can be presented in the form%
\begin{equation}
	\left.  \frac{d^{3}w\left(  t,t_{\mathrm{in}}\right)  }{k_{0}^{2}dk_{0}%
		d\Omega}\right\vert _{\underline{\mathbf{v}}_{\perp}=0}=\frac{1}{2\varepsilon
	}\left(  \frac{qc}{\pi}\right)  ^{2}\sin^{2}\theta\,\tanh\eta
	\,\operatorname{Re}I_{0}^{\left(  2\right)  }\left(  \theta;t,t_{\mathrm{in}%
	}\right)  e^{-i\Lambda\left(  \sinh\eta-\cos\theta\cosh\eta\right)  }\,.
	\label{n05}%
\end{equation}
On Fig. \ref{Fig3}, we illustrate the spectral-angular distribution
(\ref{s3.21}) as a function of $k_{x}$ (horizontal axes), $k_{\parallel}$
(vertical axes) for some values of $t$. In these pictures, we set
$c/\varepsilon=0.1$, $\left(  qc/\pi\right)  ^{2}/(2\varepsilon)=1/(5\pi^{2})$
for simplicity.%

For large $\Lambda$, we may use the asymptotic representation for
$I_{0}^{\left(  2\right)  }\left(  \theta;t,t_{\mathrm{in}}\right)  $ in
(\ref{n03}) to show that%
\begin{equation}
	\left.  \frac{d^{3}w\left(  t,t_{\mathrm{in}}\right)  }{k_{0}^{2}dk_{0}%
		d\Omega}\right\vert _{\underline{\mathbf{v}}_{\perp}=0}=\frac{q^{2}c}{2\pi
		^{2}k_{0}}\sin^{2}\theta\frac{\tanh\eta\,\tanh\eta_{\mathrm{in}}}{1-\tanh
		\eta_{\mathrm{in}}\,\cos\theta}\sin\Lambda\delta+O\left(  \Lambda^{-2}\right)
	\,,\ \ \Lambda\rightarrow\infty\,, \label{n06}%
\end{equation}
where $\delta=\sinh\eta-\sinh\eta_{\mathrm{in}}-\left(  \cosh\eta-\cosh
\eta_{\mathrm{in}}\right)  \cos\theta$. Lastly, one may use Eq. (\ref{f16}) to
show that the energy rate (\ref{s3.10}) in a symmetrical form reads:%
{\footnotesize
\begin{align}
	\left.  w\left(  T\right)  \right\vert _{\underline{\mathbf{v}}_{\perp
		}=\mathbf{0}}  &  =-\frac{q^{2}c^{2}}{2\varepsilon\pi^{2}}\int\frac{\left\vert
		\mathbf{k}_{\perp}\right\vert }{\left\vert \mathbf{k}\right\vert }\left[
	\tanh\eta_{+}\sin\left(  z\sinh u_{+}\right)  +\tanh\eta_{-}\sin\left(  z\sinh
	u_{-}\right)  \right]  \operatorname{Re}S_{0}\left(  z;T\right)
	d\mathbf{k}\nonumber\\
	&  +\frac{q^{2}c^{2}}{2\varepsilon\pi^{2}}\int\frac{\left\vert \mathbf{k}%
		_{\perp}\right\vert }{\left\vert \mathbf{k}\right\vert }\left[  \tanh\eta
	_{+}\cos\left(  z\sinh u_{+}\right)  +\tanh\eta_{-}\cos\left(  z\sinh
	u_{-}\right)  \right]  \operatorname{Im}S_{0}\left(  z;T\right)
	d\mathbf{k\,,} \label{s3.21}%
\end{align}
}
where $u_{\pm}=\eta_{\pm}-\xi$, $\eta_{\pm}=\mathrm{arc}\sinh\left(
\pm\varepsilon T/2+\underline{u}_{\Vert}/c\right)  $ and%
\begin{eqnarray*}
	S_{0}\left(  z;T\right)  &=&S_{0}\left(  z;+T/2,-T/2\right)  =K_{0}^{\prime
	}\left(  z;T\right)  -\frac{1}{z}\frac{k_{\parallel}}{\left\vert
		\mathbf{k}\right\vert }\dot{K}_{0}\left(  z;T\right)  \,,\\
	K_{0}^{\prime}\left(  z;T\right)  &=&K_{0}^{\prime}\left(
	z;+T/2,-T/2\right)  =\frac{i}{2}\int_{u_{-}}^{u_{+}}e^{iz\sinh u^{\prime}%
	}\sinh u^{\prime}du^{\prime}\,,\\
	\dot{K}_{0}\left(  z;T\right)  &=&\dot{K}_{0}\left(  z;+T/2,-T/2\right)
	=\left\{
	\begin{array}
		[c]{cl}%
		\left(  e^{iz\sinh u_{-}}-e^{iz\sinh u_{+}}\right)  /2 & \mathrm{if}%
		\ \ T<+\infty\ ,\\
		0 & \mathrm{if\ \ }T=+\infty\ .
	\end{array}
	\right.
\end{eqnarray*}
In the large time limit, $T\rightarrow\infty$, the symmetrical energy rate
(\ref{s3.21}) coincides with the result given by Eq. (\ref{s3.15}).

\section{Discussion\label{S4}}

In this work, we address the problem of the electromagnetic radiation produced
by charge distributions in a semiclassical approach, in which the radiation
field is quantum while current densities--sources of radiation--are regarded
classically. In this formulation, quantum states of the electromagnetic field
are exact solutions of Schr\"{o}dinger's equation, whose Hamiltonian describes
both free photons as well as their interaction with external currents. On this
basis, pertinent electromagnetic quantities such as energies and energy rates
radiated by currents, are calculated with the aid of transition probabilities
between states with well-defined number of photons. This construction enables
us to introduce, rigorously, the quantum transition time as a
\textquotedblleft radiation interval\textquotedblright\ and to assess its role
in radiation problems. More specifically, assuming the vacuum as the initial
state, we calculated time-dependent one-photon, multi-photon, total
electromagnetic energies, and the rate at which the radiation is emitted from
the source; the latter, in particular, presented for the first time. It must
be noted that time is intrinsically absent in classical electrodynamics, as
the electromagnetic energy and rate are usually derived through Poynting's and
Parseval's theorem. We discovered that our formulas for the total energy and
rate are compatible with the corresponding classical results in the limit
where the quantum transition interval tends to infinity. Moreover, all
quantities obtained in the semiclassical approach are valid to any current
distribution as well as the nature of the external force field responsible for
their acceleration.

To illustrate the use of the semiclassical approach, we study a simple yet
intriguing physical system: the pointlike charged particle accelerated by a
constant and uniform electric field. We present a detailed consideration of
the total energy and rate radiated by the particle, both in cases where it
performs a general trajectory in the space as when it moves parallel to the
field. Our expression for the total energy is time-dependent and coincides
with the classical result in the large-time limit. We derive an asymptotic
approximation for the angular-spectral distribution of the total energy and
represent it graphically, considering the external field fixed but varying the
radiation interval. Our results show a typical radiation pattern of linearly
accelerated charged particles: the pattern has lobes tipped toward the
direction of motion. Similar patterns are also seen in the pictures for the
energy emission rate.

Besides electromagnetic energies, we derived time-dependent expressions for
the total energy rate radiated by the particle. After integrating it over the
angles and momentum, we discovered that our result differs from Larmor's by a
factor of $1/3$. This discrepancy is due to a difference in the definition of
energy rate. In classical electrodynamics, the rate is defined through
Parseval's theorem, while in the semiclassical formulation, the rate is
defined as the time derivative of the energy. As discussed above, the absence
of this factor was pointed out before by Nikishov and Ritus in \cite{NikRit69}
in the context of the classical theory by effectively differentiating the
total energy with respect to time.

We conclude this work by emphasizing that the semiclassical approach offers an
alternative description of physical systems interacting with background
fields. Despite being an approximation compared to QED, the semiclassical
formulation exactly incorporates the quantum character of the electromagnetic
field. For this reason, this theory allows extracting information about
electromagnetic properties stemming from the interaction between radiation and
matter beyond the reach of classical electrodynamics. We hope that the ideas
discussed in this work might help study more complex systems, e.g., involving
complicated external fields whose analytical forms do not admit to solving
relativistic wave equations exactly, thus precluding the use of Furry representation.

\section{Acknowledgments}

The work of T. C. Adorno was supported by the XJTLU Research Development
Funding, award no. RDF-21-02-056 (sections \ref{S1},\ref{S3}) and the
work of A. I. Breev and D. M. Gitman was supported by Russian Science
Foundation, grant no. 19-12-00042 (sections \ref{S2},\ref{S4}). D. M. Gitman
thanks CNPq for permanent support.

\end{document}